\documentclass[10pt]{article}
\usepackage{lmodern}
\usepackage{array}
\usepackage{graphicx}
\usepackage{multirow}
\usepackage{tabularx}
\usepackage{lscape}
\usepackage{rotating}
\usepackage{multirow}
\usepackage{fullpage}
\usepackage{anysize}
\marginsize{1.7cm}{1.7cm}{1.6cm}{1.5cm}
\usepackage{filecontents}
\usepackage{tablefootnote}
\usepackage{longtable}

\let\OLDthebibliography\thebibliography
\renewcommand\thebibliography[1]{
  \OLDthebibliography{#1}
  \setlength{\parskip}{0pt}
  \setlength{\itemsep}{0pt plus 0.3ex}
}

\usepackage[boxed,ruled,vlined,linesnumbered]{algorithm2e}
\usepackage{url}
\usepackage{framed}
\newcommand*\wrapletters[1]{\wr@pletters#1\@nil}
\def\wr@pletters#1#2\@nil{#1\allowbreak\if&#2&\else\wr@pletters#2\@nil\fi}
\usepackage{caption}
\usepackage{subcaption}
\usepackage[none]{hyphenat}
\usepackage{times}
\usepackage{amsmath}
\SetKwHangingKw{Local}{Local}
\SetKwHangingKw{Global}{Global}
\SetKwHangingKw{Constant}{Constant}
\SetKwHangingKw{Input}{Input}
\SetKwHangingKw{Alias}{Alias}
\SetKwHangingKw{Output}{Output}
\SetKwHangingKw{SideEffect}{Side effects}
\SetKwHangingKw{InternalEvent}{Internal event}
\SetKwHangingKw{External}{External}
\SetKwHangingKw{ExternalEvent}{External event}
\SetKwHangingKw{Precondition}{Precondition :}
\SetKwHangingKw{Postcondition}{Postcondition :}
\SetKwHangingKw{Upon}{Upon}
\SetKwHangingKw{DoForever}{Do Forever}
\SetKwHangingKw{PVab}{Persistent variables:}

\usepackage{theorem}
\usepackage{wrapfig}
\usepackage{amssymb}
\usepackage{xspace}

\newcommand{\qed}{\hfill $\blacksquare$}
\newcommand{\qedClaim}{\hfill \ensuremath{\Box}}

\newcommand{\B}{\vspace*{-\smallskipamount}}
\newcommand{\BB}{\vspace*{-\medskipamount}}
\newcommand{\BBB}{\vspace*{-\bigskipamount}}

\newcommand{\remove}[1]{}

\newcommand*{\AffFont}{%
              \usefont{\encodingdefault}{\rmdefault}{} {n}%
              \fontsize{10}{-5}%
              \selectfont}

\usepackage[T1]{fontenc}
\usepackage[utf8]{inputenc}
\usepackage{authblk}

\usepackage{setspace}
\usepackage{lineno}

\usepackage{csquotes}
\usepackage{enumitem}
\usepackage{tablefootnote}
\usepackage{longtable}

\newcommand{\Keywords}[1]{\par\noindent
{\small{\em Keywords\/}: #1}}
\usepackage{multirow}
\usepackage{hhline}

\begin{document}
%\linenumbers
\title{A Survey on {5G}: {The} Next Generation of Mobile Communication\thanks{Accepted in Elsevier Physical Communication.}}
%\date{}
\author[1]{Nisha Panwar}
\author[1]{Shantanu Sharma}
\author[2]{Awadhesh Kumar Singh}
\affil[1]{\AffFont Department of Computer Science, Ben-Gurion University, Israel. \texttt{\{panwar, sharmas\}@cs.bgu.ac.il}.}
\affil[2]{\AffFont Department of Computer Engineering, National Institute of Technology, Kurukshetra, India. \texttt{aksingh@nitkkr.ac.in}}

\maketitle

\begin{abstract}
\label{abs:abstrat}
The rapidly increasing number of mobile devices, voluminous data, and higher data rate are pushing to rethink the current generation of the cellular mobile communication. The next or fifth generation (5G) cellular networks are expected to meet high-end requirements. The 5G networks are broadly characterized by three unique features: ubiquitous connectivity, extremely low latency, and very high-speed data transfer. The 5G networks would provide novel architectures and technologies beyond state-of-the-art architectures and technologies. In this paper, our intent is to find an answer to the question: ``\textit{what will be done by 5G and how}?'' We investigate and discuss serious limitations of the fourth generation (4G) cellular networks and corresponding new features of 5G networks. We identify challenges in 5G networks, new technologies for 5G networks, and present a comparative study of the proposed architectures that can be categorized on the basis of energy-efficiency, network hierarchy, and network types. Interestingly, the implementation issues, \textit{e}.\textit{g}., interference, QoS, handoff, security-privacy, channel access, and load balancing, hugely effect the realization of 5G networks. Furthermore, our illustrations highlight the feasibility of these models through an evaluation of existing real-experiments and testbeds.
\end{abstract}

\Keywords{Cloud radio access networks; cognitive radio networks; D2D communication; dense deployment; multi-tier heterogeneous network; privacy; security; tactile Internet.}

\setcounter{page}{1}
%===========================================================================================================================
%===========================================================================================================================
% Section 1                 Section 1               Section 1               Section 1               Section 1
%===========================================================================================================================
%===========================================================================================================================
\section{Introduction}
\label{section:introduction}
The evolution of the cellular network generations is influenced primarily by continuous growth in wireless user devices, data usage, and the need for a better quality of experience (QoE). More than 50 billion connected devices are expected to utilize the cellular network services by the end of the year 2020~\cite{ericsson_50b}, which would result in a tremendous increase in data traffic, as compared to the year 2014~\cite{ericsson_mobility}. However, state-of-the-art solutions are not sufficient for the challenges mentioned above. In short, the increase of 3D (`D'evice, `D'ata, and `D'ata transfer rate) encourages the development of 5G networks.

Specifically, the fifth generation (5G) of the cellular networks will highlight and address the following three broad views: (\textit{i}) user-centric (by providing 24$\times$7 device connectivity, uninterrupted communication services, and a smooth consumer experience), (\textit{ii}) service-provider-centric (by providing a connected intelligent transportation systems, road-side service units, sensors, and mission critical monitoring/tracking services), and (\textit{iii}) network-operator-centric (by providing an energy-efficient, scalable, low-cost, uniformly-monitored, programmable, and secure communication infrastructure). Therefore, 5G networks are perceived to realize the three main features as below:
\begin{itemize}[noitemsep,leftmargin=.5cm]
  \item \textit{Ubiquitous connectivity}: In future, many types of devices will connect ubiquitously and provide an uninterrupted user experience. In fact, the user-centric view will be realized by ubiquitous connectivity.

  \item \textit{Zero latency}: The 5G networks will support life-critical systems, real-time applications, and services with zero delay tolerance. Hence, it is envisioned that 5G networks will realize zero latency, \textit{i}.\textit{e}., extremely low latency of the order of 1 millisecond~\cite{nokia,ti2014}. In fact, the service-provider-centric view will be realized by the zero latency.

  \item \textit{High-speed Gigabit connection}: The zero latency property could be achieved using a high-speed connection for fast data transmission and reception, which will be of the order of Gigabits per second to users and machines~\cite{nokia}.
\end{itemize}

A few more \textit{key features of 5G networks} are enlisted and compared to the fourth generation (4G) of the cellular networks, as below~\cite{European_Commission-5G,gsma_intelligence,metis}: (\textit{i}) 10-$100x$ number of connected devices, (\textit{ii}) 1000$x$ higher mobile data volume per area, (\textit{iii}) 10-$100x$ higher data rate, (\textit{iv}) 1 millisecond latency, (\textit{v}) 99.99\% availability, (\textit{vi}) 100\% coverage, (\textit{vii}) $\frac{x}{10}$ energy consumption as compared to the year 2010, (\textit{viii}) real-time information processing and transmission, (\textit{ix}) $\frac{x}{5}$ network management operation expenses, and (\textit{x}) seamless integration of the current wireless technologies.

\begin{wrapfigure}{r}{9cm}
\BBB
\begin{center}
\includegraphics[scale=0.45]{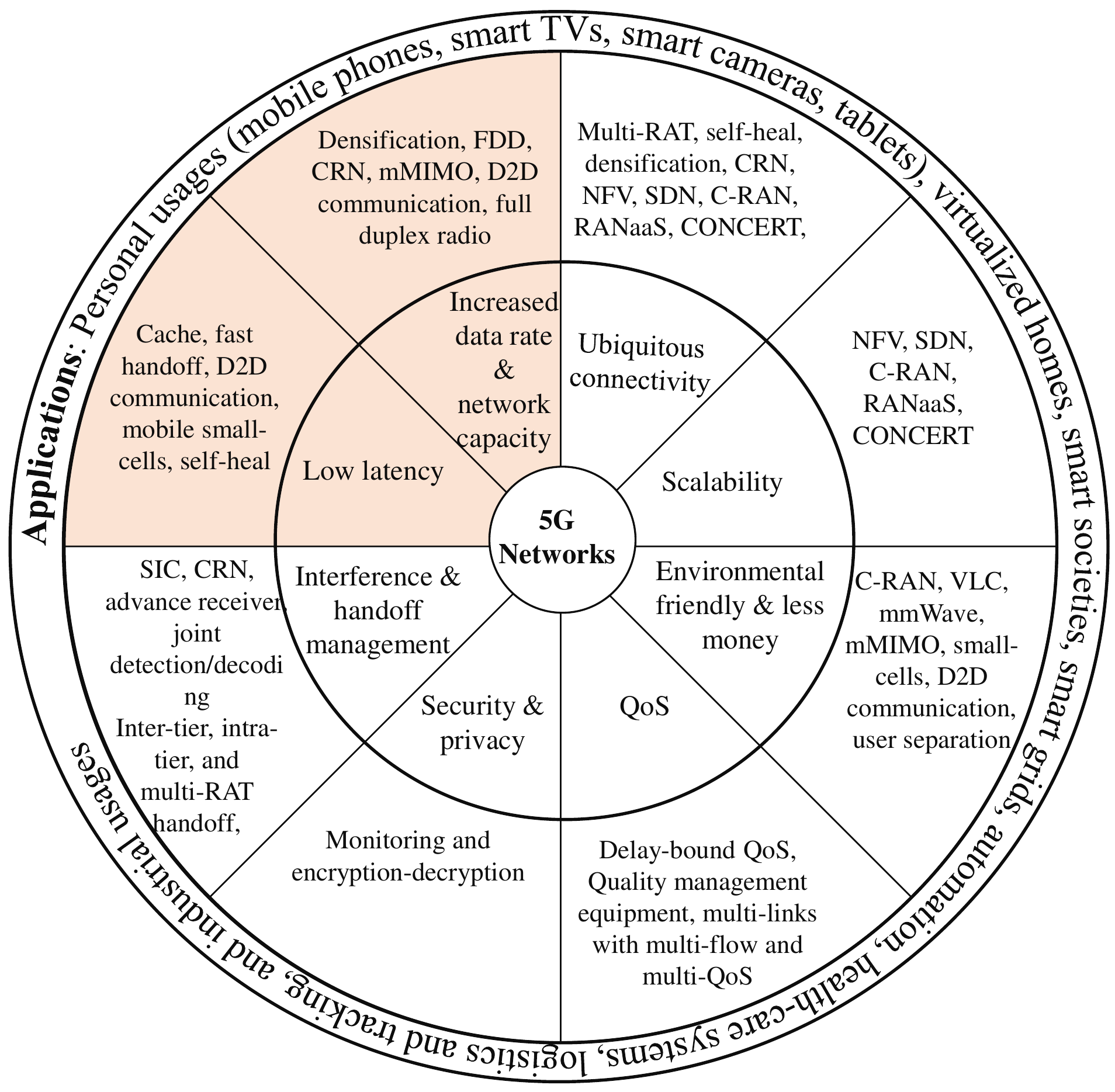}
\end{center}
\BBB
\caption{Requirements and proposed solutions for the development of 5G networks. The inner, middle, and outermost layers present requirements, solutions, and applications of 5G networks, respectively. Two colored wedges highlight primary features of 5G networks.}
\B
\label{fig:circle}
\end{wrapfigure}

The revolutionary scope and the consequent advantages of the envisioned 5G networks, therefore, demand new architectures, methodologies, and technologies (see Figure~\ref{fig:circle}), \textit{e}.\textit{g}., energy-efficient heterogeneous frameworks, cloud-based communication (software-defined networks (SDN) and network function virtualization (NFV)), full duplex radio, self-interference cancellation (SIC), device-to-device (D2D) communications, machine-to-machine (M2M) communications, access protocols, cheap devices, cognitive networks (for accessing licensed, unlicensed, and shared frequency bands), dense-deployment, security-privacy protocols for communication and data transfer, backhaul connections, massive multiple-input and multiple-output (mMIMO), multi-radio access technology (RAT) architectures, and technologies for working on millimeter wave (mmWave) 30–300 GHz. Interestingly, the 5G networks will not be a mere enhancement of 4G networks in terms of additional capacity; they will encompass a system architecture visualization, conceptualization, and redesigning at every communication layer~\cite{DBLP:journals/network/GeCGH14}.

Several industries, \textit{e}.\textit{g}., Alcatel-Lucent~\cite{Alcatel-Lucent}, DOCOMO~\cite{docomo}, GSMA Intelligence~\cite{gsma_intelligence}, Huawei~\cite{Huawei}, Nokia Siemens Networks~\cite{nokia}, Qualcomm~\cite{Qualcomm}, Samsung~\cite{samsung}, Vodafone\footnote{\url{http://www.surrey.ac.uk/5gic/research}}, the European Commission supported 5G Infrastructure Public Private Partnership (5GPPP)~\cite{European_Commission-5G}, and Mobile and Wireless Communications Enablers for the Twenty-Twenty Information Society (METIS)~\cite{metis}, are brainstorming with the development of 5G networks. Currently, the industry standards are yet to be evolved about the expected designs and architectures for 5G networks.

\begin{figure}
\centering
\includegraphics[scale=0.44]{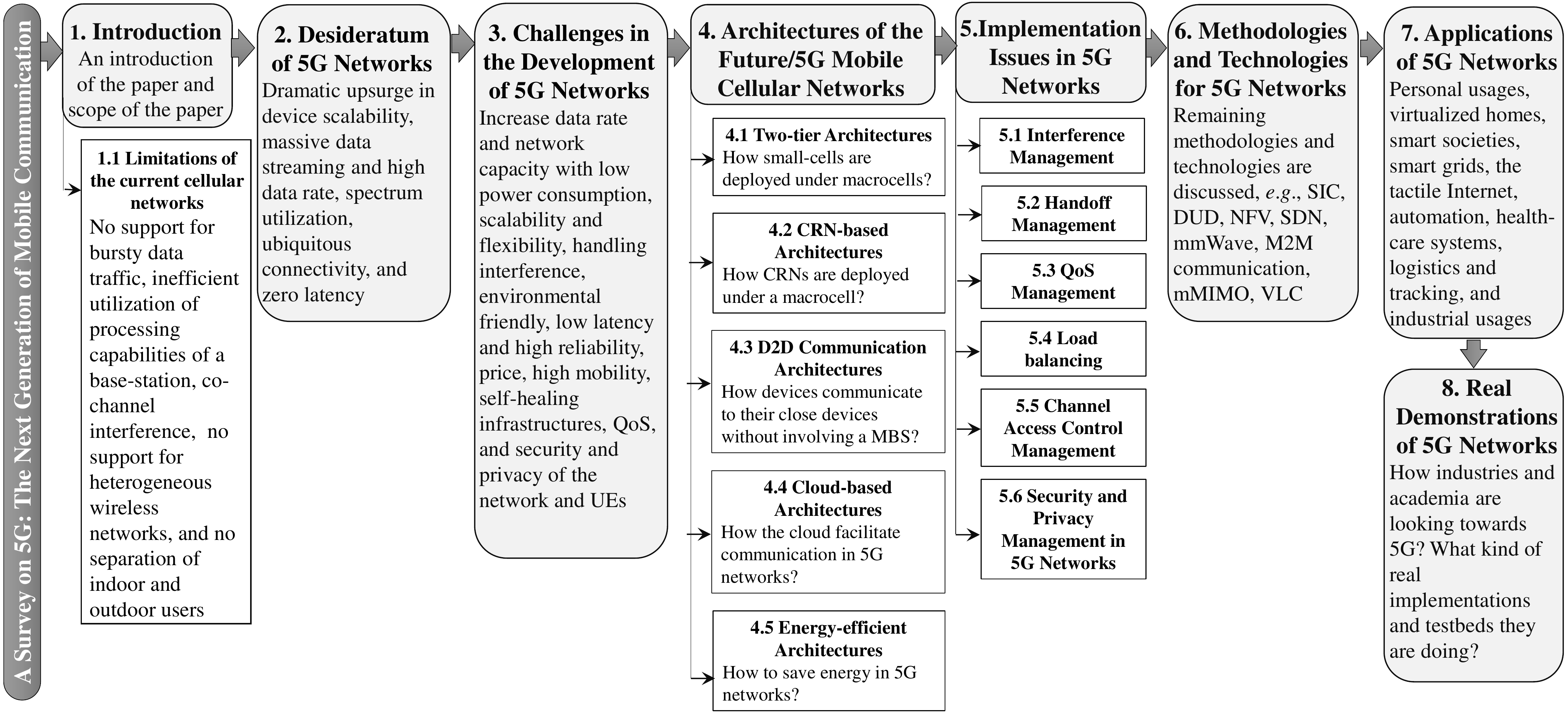}
\BBB
\caption{Schematic map of the paper.}
\label{fig:Outline of the paper}
\end{figure}

\smallskip\noindent\textbf{Scope of the paper.} In this paper, we will review the vision of the 5G networks, advantages, applications, proposed architectures, implementation issues, real demonstrations, and testbeds. The outline of the paper is provided in Figure~\ref{fig:Outline of the paper}. In Section~\ref{section:Promises of 5G}, we will elaborate the vision of 5G networks. Section~\ref{section:Challenges in the Development of 5G} presents challenges in the development of 5G networks. Section~\ref{sec:Architectures of the Future 5G Mobile Cellular Networks} addresses the currently proposed architectures for 5G networks, \textit{e}.\textit{g}., multi-tier, cognitive radio based, cloud-based, device proximity based, and energy-efficient architectures. Section~\ref{sec:Management Issues in 5G Networks} presents issues regarding interference, handoff, quality of services, load balancing, channel access, and security-privacy of the network. Sections~\ref{sec:Methodology and Technology for 5G Networks},~\ref{sec:Applications of 5G Networks}, and~\ref{sec:Real Demonstrations and Test-beds for 5G Networks} present several methodologies and technologies involved in 5G networks, applications of 5G networks, and real demonstrations and testbeds of 5G networks, respectively.

We would like to emphasize that there do exist some review works on 5G networks by Andrews et al.~\cite{DBLP:journals/jsac/AndrewsBCHLSZ14}, Chávez-Santiago et al.~\cite{5g_review1}, and Gavrilovska et al.~\cite{5g_review2}, to the best of our knowledge. However, our perspective about 5G networks is different, as we deal with a variety of architectures and discuss several implementation affairs, technologies in 5G networks along with applications and real-testbed demonstrations. In addition, we intentionally avoid an mmWave oriented discussion in this paper, unlike the current work~\cite{DBLP:journals/jsac/AndrewsBCHLSZ14,5g_review1,5g_review2}.

We encourage our readers to see an overview about the generations of the cellular networks (see Table~\ref{table:The generation of cellular networks}) and the crucial limitations of current cellular networks in the next section.

\begin{table}[h]
%\small
\begin{center}
\bgroup
\def\arraystretch{1.1}
\centering
\begin{tabular}{|l|l|p{8cm}|p{5cm}|}
\hline

\textbf{Generations} & \textbf{Year} & \textbf{Features} & \textbf{Limitations} \\\hline\hline

1G & 1980s & Analog signals for voice only communications & Very less security \\\hline

2G & 1990s & Digital signals, voice communications, and text messaging & Very less support for the Internet \\\hline

3G & 1998-99 & Voice communications, wireless mobile and fixed Internet access, video calls, and mobile television (TV) & Less support for high-speed Internet \\\hline

4G & 2008-09 & Higher data rate (hundreds of megabits per second) & No support for 50 billion ubiquitous connected devices \\\hline

5G & 2020 & Mentioned in Section~\ref{section:introduction} & \\\hline

\end{tabular}
\egroup
\caption{The generations of the cellular networks.}
\label{table:The generation of cellular networks}
\end{center}
\end{table}

\subsection{Limitations of the Conventional Cellular Systems}
\label{subsec:Limitations of the Conventional Cellular Systems}
The 4G networks are not substantial enough to support massively connected devices with low latency and significant spectral efficiency, which will be crucial in the future communication and computing. In this section, we discuss a few crucial aspects in which conventional cellular networks lag far behind, thereby motivating the evolution of 5G networks.

\smallskip\noindent\textbf{No support for bursty data traffic.} There are several mobile applications that send heartbeat messages to their servers and occasionally request for very high data transfer rate for a very short duration. Such types of data transmission consume more battery life of (mobile) user equipments (UEs) with increasing bursty data in the network, and hence, may crash the core network~\cite{DBLP:journals/network/ZhouZLZCNZ14}. However, only one type of signaling/control mechanism is designed for all types of the traffic in the current networks, creating high overhead for bursty traffic~\cite{DBLP:journals/cm/IRHXLP14,DBLP:conf/globecom/AuZNYBVMZ14}.

\smallskip\noindent\textbf{Inefficient utilization of processing capabilities of a base-station.} In the current cellular networks, the processing power of a base-station (BS) can only be used by its associated UEs, and they are designed to support peak time traffic. However, the processing power of a BS can be shared across a large geographical area when it is lightly loaded. For example: (\textit{i}) during the day, BSs in business areas are over-subscribed, while BSs in residential areas are almost idle, and vice versa~\cite{DBLP:journals/network/WuZHW15}, and (\textit{ii}) BSs in residential areas are overloaded in weekends or holidays while BSs in business areas are almost idle~\cite{DBLP:journals/twc/OhSK13}. However, the almost idle BSs consume an identical amount of power as over-subscribed BSs; hence, the overall cost of the network increases.

\smallskip\noindent\textbf{Co-channel interference.} A typical cellular network uses two separate channels, one as a transmission path from a UE to a BS, called uplink (UL), and the reverse path, called downlink (DL). The allocation of two different channels for a UE is not an efficient utilization of the frequency band. However, if both the channels operate at an identical frequency, \textit{i}.\textit{e}., a full duplex wireless radio~\cite{DBLP:conf/sigcomm/BharadiaMK13}, then a high level of co-channel interference (the interference between the signals using an identical frequency) in UL and DL channels is a major issue in 4G networks~\cite{DBLP:journals/cm/NamBLK14}. It also prevents the \textit{network densification}, \textit{i}.\textit{e}., the deployment of many BSs in a geographical area.

\smallskip\noindent\textbf{No support for heterogeneous wireless networks.} The heterogeneous wireless networks (HetNets) are composed of wireless networks with diverse access technologies, \textit{e}.\textit{g}., the third generation (3G), 4G, wireless local area networks (WLAN), WiFi, and Bluetooth. The HetNets are already standardized in 4G; however, the basic architecture was not intended to support them. Furthermore, the current cellular networks allow a UE to have a DL channel and a UL channel must be associated with a single BS that prevents the maximum utilization of HetNets. In HetNets, a UE may select its UL and DL channels from two different BSs belonging to two different wireless networks for performance improvement~\cite{DBLP:journals/cm/BoccardiHLMP14,DBLP:conf/globecom/ElshaerBDI14}.

\smallskip\noindent\textbf{No separation of indoor and outdoor users.} The current cellular networks have a single BS installed preferably near the center of the cell and interacts with all the UEs irrespective of the indoor or outdoor location of the UEs; while UEs stay indoors and outdoors for about 80\% and 20\% of the time, respectively. Furthermore, the communication between an indoor UE and an outside BS is not efficient in terms of data transfer rate, spectral efficiency, and energy-efficiency, due to the attenuation of signals passing through walls~\cite{DBLP:journals/cm/WangHGYYYAHFH14}.

\smallskip\noindent\textbf{Latency.} When a UE receives an access to the best candidate BS, it takes several hundreds of milliseconds in the current cellular networks~\cite{DBLP:journals/cm/ZhangZWZIPLC15}, and hence, they are unable to support the zero latency property.

%==================================================================================================
%==================================================================================================
%Need of 5G                   Need of 5G                 Need of 5G            Need of 5G
%==================================================================================================
%==================================================================================================
\section{Desideratum of 5G Networks}
\label{section:Promises of 5G}
A growing number of UEs and the corresponding surge in the bandwidth requirement for the huge amount of data transmission certainly necessitate the novel enhancement to the current technology. In this section, we highlight requirements of the future 5G networks.

\smallskip\noindent\textbf{Dramatic upsurge in device scalability.} A rapid growth of smart phones, gaming consoles, high-resolution TVs, cameras, home appliances, laptops, connected transportation systems, video surveillance systems, robots, sensors, and wearable devices (watches and glasses) is expected to continue exponentially in the near future. Therefore, the 5G networks are perceived to support massively connected devices~\cite{DBLP:journals/cm/WangHGYYYAHFH14,ericsson_50b,DBLP:journals/cm/AgyapongISKB14}.

\smallskip\noindent\textbf{Massive data streaming and high data rate.} A vast growth in a number of wireless devices will of course result in a higher amount of data trading (\textit{e}.\textit{g}., videos, audio, Web browsing, social-media data, gaming, real-time signals, photos, bursty data, and multimedia) that will be 100-times more as compared to the year 2014 and would overburden the current network. Thus, it is mandatory to have matching data transfer capabilities in terms of new architectures, methods, technologies, and data distribution of indoor and outdoor users~\cite{DBLP:journals/wc/HossainRTA14,DBLP:journals/cm/AgyapongISKB14,DBLP:journals/cm/HongWWS14}.

\smallskip\noindent\textbf{Spectrum utilization.} The two different channels (one for a UL and another for a DL) seem redundant from the point of view of the spectrum utilization~\cite{DBLP:journals/cm/HongBCJMKL14}. In addition, the currently allocated spectrums have their significant portions under-utilized~\cite{url1}. Hence, it is necessary to develop an access control method that can enhance the spectrum utilization. Furthermore, the spectrum utilization and efficiency have already been stretched to the maximum. It definitely requires spectrum broadening (above 3 GHz) along with novel spectrum utilization techniques~\cite{DBLP:journals/cm/ChenZ14}.

\smallskip\noindent\textbf{Ubiquitous connectivity.} Ubiquitous connectivity requires UEs to support a variety of radios, RATs, and bands due to the global non-identical operating bands. In addition, the major market split between time division duplex (\textit{e}.\textit{g}., India and China) versus frequency division duplex (\textit{e}.\textit{g}., US and Europe) so that UEs are required to support different duplex options. Hence, 5G networks are envisioned for \textit{seamless connectivity} of UEs over HetNets~\cite{outlook7}.

\smallskip\noindent\textbf{Zero latency.} The future mobile cellular networks are expected to assist numerous real-time applications, the tactile Internet~\cite{ti2014,DBLP:journals/cm/FettweisA14}, and services with varying levels of quality of service (QoS) (in terms of bandwidth, latency, jitter, packet loss, and packet delay) and QoE (in terms of users' and network-providers' service satisfaction versus feedback). Hence, 5G networks are envisioned to realize real-time and delay-bound services with the optimal QoS and QoE experiences~\cite{DBLP:journals/cm/AgyapongISKB14,DBLP:journals/cm/NamBLK14}.

%==================================================================================================
%==================================================================================================
%Challenges in 5G             Challenges in 5G              Challenges in 5G            Challenges in 5G
%==================================================================================================
%==================================================================================================
\section{Challenges in the Development of 5G Networks}
\label{section:Challenges in the Development of 5G}

The vision of 5G networks is not trivial to achieve. There are several challenges (some of the following challenges are shown in Figure~\ref{fig:circle} with their proposed solutions) to be handled in that context, as mentioned below:

\smallskip\noindent\textbf{Data rate and network capacity expansion with energy optimization.}
The deployment of more BSs in a geographical area, use of the higher frequency bands, and link improvement might support the network capacity expansion, billions of UEs, high data rate, high volume of data, and efficient backhaul data transfer to the core network. However, the implementation of these solutions is a cumbersome task in terms of economy and energy intake. Hence, the network capacity is required to be significantly increased, keeping the energy consumption and cost under strict control.

\textit{Proposed solutions}: Network densification or small-cell deployment~\cite{DBLP:journals/cm/AgyapongISKB14,DBLP:journals/cm/BhushanLMGBDSPG14,DBLP:journals/cm/WangHGYYYAHFH14} (Section~\ref{subsec:Two-tier architectures}), cognitive radio networks (CRNs)~\cite{Akyildiz:2009:CCR:1508337.1508834} (Section~\ref{subsec:Cognitive Radio Network based Architectures}), mMIMO~\cite{DBLP:journals/cm/LarssonETM14,DBLP:journals/jstsp/LuLSAZ14,DBLP:journals/cm/NamNSL0KL13} (Section~\ref{sec:Methodology and Technology for 5G Networks}), network offload using D2D communication~\cite{DBLP:conf/globecom/ChenZZM14,DBLP:journals/cm/TehraniUY14,DBLP:journals/cm/WeiHQW14} (Section~\ref{subsec:Device-to-Device Communication Architectures}), efficient backhaul networks~\cite{DBLP:journals/network/GeCGH14,DBLP:journals/wc/NiCWL14} (Section~\ref{subsubsec:Backhaul data transfer from small-cells}), energy-efficient architectures~\cite{DBLP:journals/cm/HuQ14,DBLP:journals/telsys/MavromoustakisB15} (Section~\ref{subsec:Energy-Efficient Architectures for 5G Networks}), full duplex radios~\cite{DBLP:conf/sigcomm/BharadiaMK13} (Section~\ref{sec:Methodology and Technology for 5G Networks}), NFV, and SDN based architectures~\cite{china,DBLP:journals/wc/LiuZZCN14,DBLP:journals/cm/RostBDGLMSW14,DBLP:journals/cm/YaziciKS14} (Section~\ref{sec:Methodology and Technology for 5G Networks}).

\smallskip\noindent\textbf{Scalability and flexibility.} These are the most prominent features of the future mobile communication. The future cellular infrastructures and methodologies must be designed to work in HetNets. Moreover, a vast number of potential users might request simultaneously for a set of services. Therefore, 5G networks must be powerful enough to support a scalable user demand across the coverage area~\cite{DBLP:journals/wc/LiuZZCN14,DBLP:journals/cm/OsseiranBBKMMQSSHTUTF14}.

\textit{Proposed solutions}: NFV- and SDN-based architectures~\cite{china,DBLP:journals/wc/LiuZZCN14,DBLP:journals/cm/RostBDGLMSW14,DBLP:journals/cm/YaziciKS14} (Section~\ref{sec:Methodology and Technology for 5G Networks}).

\smallskip\noindent\textbf{Single channel for both UL and DL.} A \textit{full duplex wireless radio}~\cite{DBLP:conf/sigcomm/BharadiaMK13} uses only a single channel for transmitting and receiving signals at identical time and frequency. Thus, a full duplex system achieves an identical performance as having different UL and DL channels, and hence, increases link capacity, saves the spectrum, and cost. However, the implementation of full duplex systems is not trivial, because now a radio has to use sophisticated protocols for the physical and the data link layers~\cite{DBLP:journals/cm/ZhangCLVH15}, and mechanisms to remove the effects of interference~\cite{DBLP:journals/cm/HongBCJMKL14}. The advantages of a full duplex radio in 5G networks are given in~\cite{DBLP:conf/globecom/HanIXPP14,DBLP:journals/cm/HongBCJMKL14,DBLP:journals/cm/IRHXLP14}.

\smallskip\noindent\textbf{Handling interference.} Handling interference among communicating devices is a well-known challenge in the wireless communication. Due to a growing number of UEs, technologies (\textit{e}.\textit{g}., HetNets, CRNs, full duplex, and D2D communication) and applications, the interference will also increase in 5G networks, and the state-of-the-art technique may not perform well in the future cellular networks~\cite{DBLP:journals/wc/HossainRTA14}. In 5G networks, a UE may receive interference from multiple macrocell base-stations (MBSs), various UEs, and small-cell base-stations (SBSs). Hence, it is required to develop an efficient (in terms of avoiding network overload) and reliable (in terms of perfect interference detection and decoding) interference management technique for channel allocation, power control, cell association, and load balancing.

\textit{Proposed solutions}: Self-interference cancellation~\cite{DBLP:journals/cm/IRHXLP14,DBLP:journals/cm/HongBCJMKL14}, an advance receiver with interference joint detection/decoding, and network-side interference management~\cite{DBLP:journals/cm/NamBLK14}. We will discuss these solutions in Section~\ref{subsec:Interference Management in 5G Networks}.

\smallskip\noindent\textbf{Environmentally friendly.} The current radio access network (RAN) consumes $70\%$-$80\%$ of the total power~\cite{DBLP:journals/cm/IRHXLP14,DBLP:journals/wc/WuYLL15}. The wireless technologies consume lots of energy that lead to huge CO$_2$ emission and inflate the cost. It is a serious threat to the environment~\cite{DBLP:journals/cm/WangHGYYYAHFH14}. Thus, it is required to develop energy-efficient communication systems, hardware, and technologies, thereby the ratio between the network throughput and energy consumption is equitable.

\textit{Proposed solutions}: Cloud-RAN (C-RAN)~\cite{DBLP:journals/wc/WuYLL15,DBLP:journals/cm/HuQ14}, visual light communication (VLC)~\cite{DBLP:journals/wc/WuYLL15}, mmWave~\cite{DBLP:journals/wc/WuYLL15}, separation of indoor and outdoor users~\cite{DBLP:journals/wc/WuYLL15}, joint investigation of spectral efficiency and energy-efficacy~\cite{DBLP:journals/cm/IRHXLP14,DBLP:journals/cm/HuQ14}, multi-tier architectures~\cite{DBLP:journals/cm/HuQ14}, D2D communication~\cite{DBLP:conf/globecom/ChenZZM14,DBLP:journals/cm/TehraniUY14,DBLP:journals/cm/WeiHQW14}, mMIMO architectures~\cite{DBLP:journals/cm/HuQ14}, and full duplex radios~\cite{DBLP:journals/cm/IRHXLP14}. Except the above mentioned solutions, we will discuss some special techniques/architectures in the context of energy-efficiency in 5G networks in Section~\ref{subsec:Energy-Efficient Architectures for 5G Networks}.

\smallskip\noindent\textbf{Low latency and high reliability.} Low latency and high reliability are critical in several real-time applications, \textit{e}.\textit{g}., message transmission by robots monitoring patients, life safety systems, cloud-based gaming, nuclear reactors, sensors, drones, and connected transportation systems. However, it is challenging to have extremely low latency and reliable delivery of data over a large scale network without increasing the network infrastructure cost, as it requires the development of techniques providing fast connections, quick handovers, and high data transfer rate.

\textit{Proposed solutions}: Caching methods~\cite{DBLP:journals/cm/BoccardiHLMP14,DBLP:journals/cm/WangCTKL14}, VLC, mmWave, mMIMO (Section~\ref{sec:Methodology and Technology for 5G Networks}), fast handover techniques~\cite{DBLP:journals/cm/DuanW15,DBLP:journals/corr/OrsinoAMI15,DBLP:journals/tvt/SongFY14} (Section~\ref{subsec:Handoff Management in 5G Networks}), and D2D communication (Section~\ref{subsec:Device-to-Device Communication Architectures}).

\smallskip\noindent\textbf{Network performance optimization.} The performance parameters, \textit{e}.\textit{g}., peak data rate, geographical area coverage, spectral efficiency, QoS, QoE, ease of connectivity, energy-efficiency, latency, reliability, fairness of users, and implementation complexity, are crucial for a cellular network~\cite{DBLP:journals/cm/WangHGYYYAHFH14}. Hence, a general framework for 5G networks should substantially optimize these parameters. However, there are some tradeoffs among all parameters, which further emphasize the need of a joint optimization algorithm.

\smallskip\noindent\textbf{Economical impacts.} A revolutionary change in the future mobile communication techniques would have drastic economical impacts in terms of deployment and motivation for user participation. It is critical to provide an entirely new infrastructure due to economical stretch. Therefore, the cost of deployment, maintenance, management, and operation of an infrastructure must be affordable from the perspective of governments, regulating authorities, and network operators. Also, the cost of using D2D communication should be feasible, so that devices involved in D2D communication should not charge more than using the services of a BS~\cite{DBLP:journals/cm/AgyapongISKB14,DBLP:journals/cm/FettweisA14}. Further, the projected revenue growth is much lower than the traffic growth~\cite{china}; hence, it is required to develop 5G networks in a manner that both network operators and users get honey in their hands.

%\centering
%\begin{longtable}[t]{|l|l|l|l|l|l|l|l|l|l|l|l|l|}
\begin{table}[t]
%\small
\begin{center}
 \begin{tabular}{|l|l|l|l|l|l|l|l|l|l|l|}
\hline
    Methodologies/Technologies & {\rotatebox{90}{\parbox{2.5cm}{Section}}} & {\rotatebox{90}{\parbox{2.55cm}{Increase data rate}}} & {\rotatebox{90}{\parbox{2.55cm}{Increase network capacity}}} & {\rotatebox{90}{\parbox{2.25cm}{Massive device support}}} & {\rotatebox{90}{\parbox{2.5cm}{Energy-efficient}}} & {\rotatebox{90}{\parbox{2.25cm}{Low latency}}} & {\rotatebox{90}{\parbox{2.25cm}{Economic}}} & {\rotatebox{90}{\parbox{2.25cm}{Security and privacy}}} & {\rotatebox{90}{\parbox{2.25cm}{Interference}}} & {\rotatebox{90}{\parbox{2.25cm}{Mobility support}}} \\ \hline\hline
    Small-cells & \ref{subsec:Two-tier architectures} & \checkmark & \checkmark &  & \checkmark &  & \checkmark$^P$\tablefootnote{\checkmark$^P$: Partial support} &  &  &  \\ \hline

    Mobile small-cells & \ref{subsec:Two-tier architectures} & \checkmark & \checkmark &  & \checkmark &  & \checkmark$^P$ &  &  & \checkmark \\ \hline

    CRN & \ref{subsec:Cognitive Radio Network based Architectures}  &  & \checkmark &  &  &  & \checkmark &  &  &  \\ \hline

    D2D & \ref{subsec:Device-to-Device Communication Architectures} & \checkmark$^P$ &  & \checkmark$^P$ & \checkmark &  & \checkmark$^P$ &  & \checkmark$^P$ &  \\ \hline

    C-RANs & \ref{subsec:Cloud-based Architectures} & \checkmark & \checkmark & \checkmark & \checkmark &  & \checkmark &  & \checkmark & \checkmark \\ \hline

    Full duplex radio & \ref{section:Challenges in the Development of 5G},\ref{subsec:Interference Management in 5G Networks}  &  & \checkmark &  &  &  &  &  & \checkmark &  \\ \hline

    Advance receiver & \ref{subsec:Interference Management in 5G Networks}  &  & \checkmark  &  &  &  &  &  & \checkmark &  \\ \hline

    SIC & \ref{subsec:Interference Management in 5G Networks},\ref{sec:Methodology and Technology for 5G Networks} &  & \checkmark &  &  & \checkmark$^P$ &  &  & \checkmark & \checkmark$^P$ \\ \hline

    DUD & \ref{sec:Methodology and Technology for 5G Networks} &  & \checkmark$^P$ &  & \checkmark &  &  &  &  &  \\ \hline

    mmWave & \ref{sec:Methodology and Technology for 5G Networks}  & \checkmark & \checkmark & \checkmark & \checkmark & \checkmark &  &  &  &  \\ \hline

    mMIMO & \ref{sec:Methodology and Technology for 5G Networks}  & \checkmark & \checkmark & \checkmark & \checkmark & \checkmark &  &  & \checkmark &  \\ \hline

    VLC & \ref{sec:Methodology and Technology for 5G Networks} & \checkmark &  &  & \checkmark & \checkmark & \checkmark & \checkmark &  &  \\ \hline

    CCN-based caching & \ref{sec:Methodology and Technology for 5G Networks} &    &  &  &  &  & \checkmark &  &  &  \\ \hline
    \end{tabular}
\caption{Summary of methodologies and technologies for 5G networks.}
\label{table:all_technique}
\BBB
\end{center}
%\end{longtable}
\end{table}

\smallskip\noindent\textbf{High mobility and handoff.} The 5G wireless UEs are meant for retaining an active service connection while frequently moving from one cell to another or from one RAT (\textit{e}.\textit{g}., 3G, 4G, 5G, WiFi, Bluetooth, and WLAN) to another. The mobility adaptation for the wireless services should not back-off even at a very high speed as a UE inside a moving vehicle. Moreover, during a particular interval, many UEs move from one place to another; for example, moving to offices from residential areas in the morning. As a result, 5G networks are envisioned to use the spectrum in the best manner and to cope up with pace of the device movement.

\textit{Proposed solutions}: Inter-tier, intra-tier, and multi-RATs handoff mechanisms, and a mechanism for secure handoff~\cite{DBLP:journals/cm/DuanW15,DBLP:journals/corr/OrsinoAMI15,DBLP:journals/tvt/SongFY14,DBLP:journals/cm/GiustCB15}, which we will discuss in Section~\ref{subsec:Handoff Management in 5G Networks}.

\smallskip\noindent\textbf{Self-healing infrastructures.} A self-healing infrastructure finds a \textit{failed} macrocell or small-cell (\textit{i}.\textit{e}., a cell that is unable to work because of hardware failures, software failures, or misconfigurations) with the help of neighboring cells and provides a way for communication to the affected users by adjusting the transmission power and operating channels in the neighboring cells~\cite{DBLP:journals/cm/ElSawyHK13,DBLP:journals/wc/WangZ14}. The design of a self-healing network insists on the frequent communication among cells; hence, it brings in the following challenges, as: (\textit{i}) develop an efficient algorithm that can detect and reconfigure a failed cell with insignificant communication and computational overheads in the minimal detection time, and (\textit{ii}) reconfiguration of a failed cell should not lead to degradation of nearby cells' services.

\textit{Proposed solutions}: A small-cell network with self-healing property is suggested in~\cite{DBLP:journals/wc/WangZ14}, which we will discuss in Section~\ref{subsubsec:two-tier with self-healing property}.

\smallskip\noindent\textbf{QoS.} QoS guarantee in 5G networks has inherent difficulties, \textit{e}.\textit{g}., node mobility, multi-hop communication, resource allocation, and lack of central coordination. In addition, in 5G networks, a huge amount of bursty and multimedia data, multi-RATs, and low latency bound for different applications and services are major hurdles in achieving the desired QoS. Hence, it is challenging to design fast and efficient algorithms to maintain real-time QoS without overloading a BS~\cite{DBLP:journals/network/ZhouZLZCNZ14,DBLP:journals/network/ZhangCZ14}.

\textit{Proposed solutions}: Delay-bound QoS~\cite{DBLP:journals/cm/HuQ14,DBLP:journals/network/ZhangCZ14}, intelligent equipment~\cite{DBLP:journals/network/ZhouZLZCNZ14}, and multi-link with multi-flow and multi-QoS~\cite{DBLP:conf/icc/KimM14} have been suggested, which we will discuss in Section~\ref{subsec:QoS Management in 5G Networks}.

%\smallskip\noindent\textbf{IoT.}

\smallskip\noindent\textbf{Security and privacy of the network and UEs.} The promising features of 5G networks bring in hard challenges in the design of security and privacy oriented 5G networks. For example, a huge number of new types of social (all-time connected) devices may originate several types of attacks like impersonation, denial-of-services (DoS), replay, eavesdropping, man-in-the-middle, and repudiation attacks. Also, the transfer of a huge volume of data in secure and high speed manners is critical while preventing malicious files to penetrate. In addition, the network densification needs to be secure and requires fast-secure handoff of UEs. We further highlight challenges in security and privacy of the network and UEs in Section~\ref{subsec:Security and Privacy Management in 5G Networks}.

\textit{Proposed solutions}: Physical layer security~\cite{DBLP:journals/cm/YangWGEYR15}, monitoring~\cite{DBLP:journals/wpc/Ulltveit-MoeOK11,gai2015intrusion,DBLP:journals/wpc/LiKA11}, secret adaptive frequency hopping~\cite{DBLP:journals/wpc/LiKA11}, encrypted-~\cite{DBLP:journals/network/LiuASLLS15}, and policy-based communications~\cite{policy-based}, which we will discuss in Section~\ref{subsec:Security and Privacy Management in 5G Networks}.

All the above mentioned methodologies and technologies are comparatively studied in Table~\ref{table:all_technique}.

\section{Architectures for the Future/5G Mobile Cellular Networks}
\label{sec:Architectures of the Future 5G Mobile Cellular Networks}
In this section, we elaborate on the existing architectures for 5G networks, namely multi-tier, CRN-based, D2D communication based, and the cloud-based architectures. These proposed 5G architectures will be explained in the light of relevant advantages, disadvantages, and the challenges that are yet to be resolved.
\subsection{Two-tier Architectures}
\label{subsec:Two-tier architectures}
\begin{figure}
\begin{center}
\includegraphics[scale=0.5]{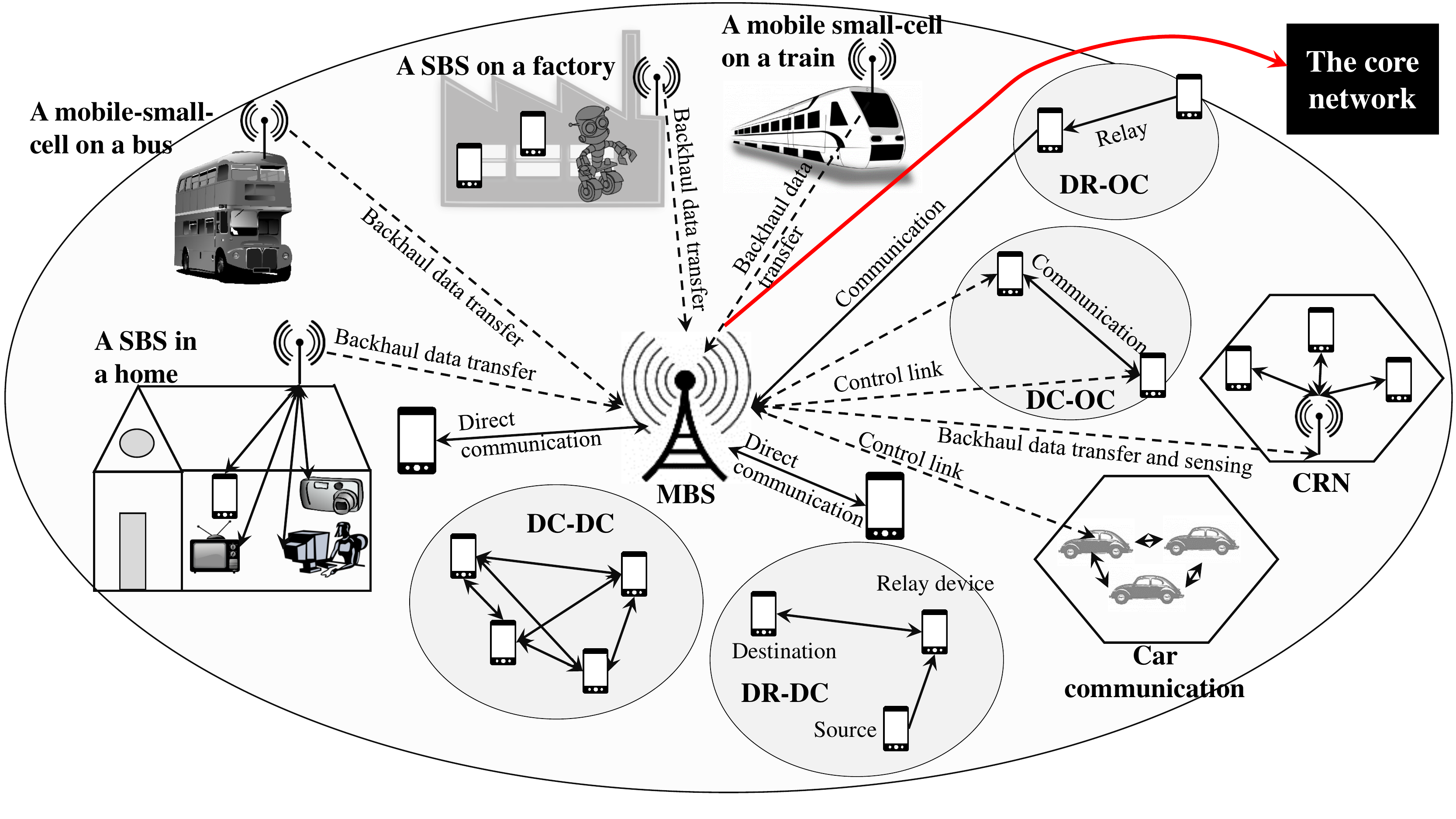}
\end{center}
\BBB
\caption{A multi-tier architecture for 5G networks with small-cells, mobile small-cells, and D2D- and CRN-based communications.}
\label{fig:Two-tier architecture for 5G networks with small-cells}
\end{figure}

Several two-tier architectures have been proposed for 5G networks, where a MBS stays in the top-tier and SBSs work under the supervision of the MBS in the lower tier. A macrocell covers all the small-cells of different types, \textit{e}.\textit{g}., femtocell, picocell, and microcell (see Table~\ref{table:Classification of cells}), and both the tiers share an identical frequency band. The small-cell enhances the coverage and services of a macrocell, and the advantages of small-cells are mentioned at the end of this section. In addition, D2D communication and CRN-based communication enhance a 2-tier architecture to a multi-tier architecture; see Figure~\ref{fig:Two-tier architecture for 5G networks with small-cells}. Note that in this section, we confine ourselves on the deployment of small-cells under the cover of a macrocell; the discussion of CRN-based and D2D communications will be carried out in Sections~\ref{subsec:Cognitive Radio Network based Architectures} and~\ref{subsec:Device-to-Device Communication Architectures}, respectively.

\begin{wraptable}{r}{7cm}
\BBB
%\begin{table}[h]
%\small
\begin{center}
\bgroup
\def\arraystretch{1}
\centering
\begin{tabular}{|l|l|l|}
\hline
\textbf{Cells} & \textbf{Range} & \textbf{Users} \\\hline
Femtocell & 10-20 meters & A few users \\\hline
Picocell & 200 meters & $20~40$ \\\hline
Microcell & 2 kilometers & $>$ 100 \\\hline
Macrocell & 30-35 kilometers & Many \\\hline
\end{tabular}
\egroup
\caption{Classification of the cells.}
\label{table:Classification of cells}
\BB
\end{center}
%\end{table}
\end{wraptable}

Wang et al.~\cite{DBLP:journals/cm/WangHGYYYAHFH14} suggested a way for separating indoor and outdoor users and using a \emph{mobile small-cell} on a train or a bus. For separating indoor and outdoor users, a MBS holds large antenna arrays with some antenna elements distributed around the macrocell and connected to the MBS using optical fibers. A SBS and large antenna arrays are deployed in each building for communicating with the MBS. All UEs inside a building can have a connection to another UE either through the SBS or by using WiFi, mmWave, or VLC. Thus, the separation of users results in less load on a MBS.

Wang et al.~\cite{DBLP:journals/cm/WangHGYYYAHFH14} also suggested to use a mobile small-cell that is located inside a vehicle to allow communication among internal UEs, while large antenna arrays are located outside the vehicle to communicate with a MBS. Thus, all the UEs inside a vehicle (or a building) appear to be a single unit with respect to the corresponding MBS, and clearly, the SBS appears as a MBS to all these UEs.

%\begin{wrapfigure}{r}{14cm}
%\BBB\BBB
%\begin{center}
%\includegraphics[scale=0.45]{pic_multi-tier.pdf}
%\end{center}
%\BBB\BB
%\caption{A multi-tier architecture for 5G networks with small-cells, mobile small-cells, and D2D- and CRN-based communications.}
%\B
%\label{fig:Two-tier architecture for 5G networks with small-cells}
%\end{wrapfigure}

In~\cite{DBLP:journals/cm/BhushanLMGBDSPG14}, a two-tier architecture is deployed as a process of \emph{network densification} that is a combination of \textit{spatial densification} (increasing the number of antennas per UE and MBS, and increasing the density of BSs) and \textit{spectral aggregation} (using higher frequency bands $>$ 3 GHz). A tradeoff between the transmission power of a macrocell and the coverage area of small-cells is presented in~\cite{DBLP:journals/cm/BhushanLMGBDSPG14}; \textit{i}.\textit{e}., on the one hand, if the transmission power of a macrocell is high, then many adjacent UEs to a small-cell may find themselves in the service area of the macrocell, and hence, it will decrease the coverage area of that small-cell. On the other hand, if the transmission power of a macrocell is low, then the coverage area of the small-cell will increase. Therefore, \emph{cell range expansion} (\textit{i}.\textit{e}., a biased handoff in the favor of small-cells) is carried out to serve more UEs by small-cells to which they are closer. Moreover, SBSs, deployed in offices or homes, can be used to serve outdoor users, \textit{e}.\textit{g}., pedestrians and low-mobility vehicles, in their neighborhoods, and the approach is called \textit{indoor-to-outdoor user service}~\cite{DBLP:journals/cm/BhushanLMGBDSPG14}.

Hossain et al.~\cite{DBLP:journals/wc/HossainRTA14} presented a multi-tier architecture consisting of several types of small-cells, relays, and D2D communication for serving users with different QoS requirements in spectrum-efficient and energy-efficient manners. Interestingly, all of these architectures consider that UEs spontaneously discover a SBS. Zhang et al.~\cite{DBLP:journals/cm/ZhangZWZIPLC15} proposed a centralized system in which a MBS assists UEs to have connections to particular SBSs, thereby interference between UEs and SBSs is reduced. However, this approach overburdens the MBS.

\smallskip\noindent\textbf{Advantage of the deployment of small-cells.}
\begin{itemize}[noitemsep,nolistsep,leftmargin=.5cm]
\item \textit{High data rate and efficient spectrum use}: The small physical separation between a SBS and UEs (served by the same SBS) leads to a higher data rate and a better indoor coverage. Also, the spectrum efficiency increases due to fewer UEs in direct communication with a MBS~\cite{DBLP:journals/wc/WangZ14}.

\item \textit{Energy saving}: The use of small-cells reduces the energy consumption of the network (by not involving MBSs) and of UEs (by allowing UEs to communicate at a shorter range with low signaling overhead)~\cite{DBLP:journals/network/GeCGH14}.

\item \textit{Money saving}: It is more economical to install a SBS without any cumbersome planning as compared to a MBS, and also the operational-management cost is much lower than the cost associated with a MBS~\cite{DBLP:journals/cm/ElSawyHK13,DBLP:journals/cm/BhushanLMGBDSPG14}.

\item The plug-and-play utility of small-cells boosts the on-demand network capacity~\cite{DBLP:journals/network/ZhouZLZCNZ14}.

\item \textit{Less congestion to a MBS}: SBSs offload UEs from a MBS so that the MBS is lightly loaded and less congested, and hence, improve the system capacity~\cite{DBLP:journals/cm/ElSawyHK13}.

\item \textit{Easy handoff}: Mobile small-cells also follow the advantages of small-cells. Moreover, they provide an attractive solution to highly mobile UEs by reducing handoff time overheads, since a mobile small-cell is capable to do the handoff on behalf of all related UEs~\cite{DBLP:journals/cm/BhushanLMGBDSPG14}.
\end{itemize}

\smallskip\noindent\textbf{Disadvantage of small-cells.} Despite numerous prominent benefits as mentioned above, there are a few realistic issues such as implementation cost and operational reliability. The small-cells indeed impose an initial cost to the infrastructure, but less than the cost associated with a MBS. Moreover, a frequent authentication is mandatory due to frequent handoff operations. Further, an active or passive (on/off) state update of any small-cell would definitely result in frequent topological updates.

\smallskip\noindent\textbf{Open issues in the deployment of 2-tier architectures using small-cells.}
\begin{itemize}[noitemsep,nolistsep,leftmargin=.5cm]
  \item \textit{Interference management}: The deployment of small-cells results in several types of interferences, as: \textit{inter-tier interference} (\textit{i}.\textit{e}., interference from a MBS to a SBS, interference from a MBS to a SBS's UEs, and interference from a SBS to a MBS's UEs), and \textit{intra-tier interference} (\textit{i}.\textit{e}., interference from a SBS to other SBSs' UEs). Hence, it is also required to develop models and algorithms to handle these interferences~\cite{DBLP:journals/cm/ElSawyHK13,DBLP:conf/globecom/DengZH14}.
  \item \textit{Backhaul data transfer}: Though we have models to transfer data from a SBS to the core network, which we will discuss next in Section~\ref{subsubsec:Backhaul data transfer from small-cells}, an extremely dense-deployment of small-cells requires a huge amount of data transfer, and certainly, requires cost efficient architectures.
\end{itemize}

\subsubsection{Backhaul data transfer from small-cells}
\label{subsubsec:Backhaul data transfer from small-cells}
Data transfer from a SBS to the core network is a challenging task, and in general, there may be three approaches to transfer (backhaul) data to the core network, as follows:
\begin{itemize}[noitemsep,nolistsep,leftmargin=.5cm]
  \item \textit{Wired optical fiber}: by establishing a wired optical fiber link from each SBS to a MBS; however, it is time-consuming and expensive.

  \item \textit{Wireless point-to-multipoint} (PTMP): by deploying a PTMP-BS at a MBS that communicates with SBSs and transfers data to the core network.

  \item \textit{Wireless point-to-point} (PTP): by using directional antennas in line-of-sight (LOS) environments; hence, it provides high capacity PTP links (same as with wired optical fibers), at a significantly lower cost.
\end{itemize}
Ge et al.~\cite{DBLP:journals/network/GeCGH14} presented two architectures based on the wireless PTMP approach. In the first (centralized) architecture, all SBSs send data using mmWave to a MBS that eventually aggregates the received data and forwards the same to the core network using fiber. In the second (distributed) architecture, all small-cells cooperatively forward data using mmWave to a specified SBS that transfers data to the core network using fiber without the explicit involvement of a MBS.

Ni et al.~\cite{DBLP:journals/wc/NiCWL14} proposed an \emph{adaptive backhaul architecture} based on the wireless PTP approach and frequency division duplex for UL and DL channels. A tree structure is used, where the root node is connected to the core network using fiber, the leaf nodes represent UEs, and other nodes represent SBSs. The data is transferred from the leaf nodes to the root node that transfers the same to the core network. The bandwidth is selected dynamically for backhaul links, as per the current bandwidth requirements, interference conditions, and the network topology. A similar approach is also presented in~\cite{DBLP:journals/cm/BhushanLMGBDSPG14}.

\subsubsection{Two-tier architectures with self-healing property}
\label{subsubsec:two-tier with self-healing property}
An automatic detection and recovery of a failed cell is an important issue in densely deployed multi-tier architectures. Wang and Zhang~\cite{DBLP:journals/wc/WangZ14} provided three approaches for designing a self-healing architecture such as below:
\begin{enumerate}[noitemsep,nolistsep,leftmargin=.5cm]
  \item \textit{Centralized approach}: A dedicated server is responsible for detecting a failed cell by measuring and analyzing abnormal behavior of users, \textit{e}.\textit{g}., received signal strengths (RSSs) at users and handoff by several users at any time from a particular cell. The server collects global information and reconfigures the failed cell. However, the approach suffers with a high communication overhead and a high computational cost.

  \item \textit{Distributed approach}: Each SBS detects failed small-cells in neighborhoods by measuring and analyzing users' handoff behavior and the neighboring small-cells' signals. Consequently, on detecting a failed cell, a SBS might increase the transmission power in order to incorporate users of the failed cell. However, the approach might not work efficiently in case that users are scattered sparsely.

  \item \textit{Local cooperative or hybrid approach}: This approach combines the benefits of both the previous approaches, and therefore, minimizes the drawback. Essentially, two steps are utilized, namely distributed trigger and cooperative detection. In the distributed trigger, each SBS collects information about users' behavior. Subsequently, a trigger message is sent to a dedicated server in case the received information thrives below a certain threshold. Hence, it does not require communication among small-cells. In the cooperative detection, the dedicated server takes the final decision based on the information received from several small-cells, resulting in a high accuracy and lower latency.
\end{enumerate}

\subsection{Cognitive Radio Network based Architectures}
\label{subsec:Cognitive Radio Network based Architectures}
A cognitive radio network (CRN)~\cite{Akyildiz:2009:CCR:1508337.1508834} is a collection of cognitive radio nodes (or processors), called secondary users (SUs) that exploit the existing spectrum opportunistically. The SUs have the \textit{LEIRA} (learning, efficiency, intelligence, reliability, and adaptively) property for scanning and operating on multiple heterogeneous channels (or frequency bands) in the absence of the licensed user(s), termed as primary user(s) (PUs), of the respective bands~\cite{DBLP:journals/ijnm/0001S14}. Each PU has a fixed bandwidth, high transmit power, and high reliability; however, the SUs work on a broad range of bandwidth with low transmit power and low reliability.

A CRN in 5G networks is used for designing multi-tier architectures, removing interference among cells, and minimizing energy consumption in the network~\cite{DBLP:journals/wc/HuangZD13,DBLP:journals/cm/ElSawyHK13,DBLP:journals/cm/WangYH13,DBLP:journals/cm/HongWWS14,DBLP:journals/wcl/LeeS14,DBLP:journals/telsys/MavromoustakisB15}.

\subsubsection{CRN-based architectures for 5G networks}
\label{subsubsec:CRN-based architectures for 5G networks}
A CRN creates a 2-tier architecture, similar to architectures discussed in Section~\ref{subsec:Two-tier architectures}; however, it is assumed that either a MBS or a SBS has cognitive properties for working on different channels.

Hong et al.~\cite{DBLP:journals/cm/HongWWS14} presented two types of CRN-based architectures for 5G networks, as: (\textit{i}) non-cooperative and (\textit{ii}) cooperative CRNs. The \textit{non-cooperative CRN} establishes a multi-RATs system, having two separate radio interfaces that operate at the licensed and temporary unoccupied channels by PUs, called cognitive channels. The SUs work only on cognitive channels and form a CRN, which overlays on the existing licensed cellular network. The two networks can be integrated in the upper layers while must be separated in the physical layer. This architecture can be used in different manners, as: (\textit{i}) the cognitive and licensed channels are used by users near a MBS and users far away from the MBS, respectively, (\textit{ii}) the cognitive and licensed channels are used for relaxed QoS and strict QoS, respectively. The \textit{cooperative CRN} uses only a licensed channel, where SUs access the channel in an opportunistic fashion when the PU of the channel is absent. This architecture can be used in different manners, as: (\textit{i}) a SBS communicates with a MBS using the licensed channel and provides service to its UEs via an opportunistic access to the licensed frequency band, (\textit{ii}) a licensed channel is used to serve UEs by a SBS and the opportunistic access to the licensed channel is used to transfer backhaul data to the MBS.

In short, the cooperative CRN~\cite{DBLP:journals/cm/HongWWS14} provides a real intuition of incorporating CRNs in 5G networks, where a SBS works as a SU, which scans activities of a macrocell and works on temporarily unoccupied frequency bands (known as \textit{spectrum holes}~\cite{Akyildiz:2006:NGS:1162469.1162470}) by a PU to provide services to their UEs with minimally disrupting macrocell activities.

A dynamic pricing model based on a game theoretic framework for cognitive small-cells is suggested in~\cite{DBLP:conf/globecom/JiangCLR14}. Since in reality SBSs' operators and MBSs' operators may not be identical and small-cells' UEs may achieve a higher data rate as compared to macrocells' UEs, the pricing model for both UEs must be different.

\subsubsection{Interference Management using CRNs}
\label{subsubsec:Interference Management using CRNs}
Huang et al.~\cite{DBLP:journals/wc/HuangZD13} provided an approach for avoiding inter-tier interference by integrating a \textit{cognitive technique} at a SBS. The cognitive technique consists of three components, as: (\textit{i}) a cognitive module, which senses the environment and collects information about spectrum holes, collision probability, QoS requirements, macrocell activities, and channel gains, (\textit{ii}) a cognitive engine, which analyzes and stores the collected information for estimating available resources, and (\textit{iii}) a self-configuration module, which uses the stored information for dynamically optimizing several parameters for efficient handoff, interference, and power management. Further, the channel allocation to a small-cell is done in a manner to avoid inter-tier and intra-tier interferences, based on Gale-Shapley~\cite{gale1962college} spectrum sharing scheme, which avoids collisions by not assigning an identical channel to neighboring small-cells.

Wang et al.~\cite{DBLP:journals/cm/WangYH13} suggested an approach for mitigating inter-tier interference based on spectrum sensing, spectrum sharing, and cognitive relay, where links between a MBS and its UEs are considered as PUs and links between a SBS and its UEs are considered as SUs. Cognitive techniques are used for detecting interference from a MBS to a SBS and vice versa, and a path loss estimation algorithm is provided for detecting interference from a small-cell's UEs to a macrocell's UEs. After detecting inter-tier interference, a small-cell shares spectrum with a macrocell using either \textit{overlay spectrum sharing scheme} (\textit{i}.\textit{e}., SUs utilize unoccupied channels, and it is applicable when a MBS and a SBS's UEs are very close or no interference is required by a macrocell's UEs) or \textit{underlay spectrum sharing scheme} (\textit{i}.\textit{e}., SUs and PUs transmit on an identical channel while restricting transmit power of SUs, and hence, resulting in a higher spectrum utilization).

Note that a CRN can be used to support D2D communication and mitigate interferences caused by D2D communication, which we will see in Section~\ref{subsec:Device-to-Device Communication Architectures}.

\smallskip\noindent\textbf{Advantages of CRNs in 5G networks.}
\begin{itemize}[noitemsep,nolistsep,leftmargin=.5cm]
  \item \textit{Minimizing interference}: By implementing a CRN at small-cells, cognitive small-cells can avoid interference very efficiently by not selecting identical channels as the channels of neighboring small-cells.

%  \item \textit{Cover large geographical areas}: The deployment of a cognitive radio networks over spectrum holes covers large geographical areas and provides low cost services to UEs.

  \item\textit{Increase network capacity}: The spectrum holes can be exploited for supporting a higher data transfer rate and enhancing bandwidth utilization~\cite{DBLP:journals/telsys/MavromoustakisB15}.
\end{itemize}

\smallskip\noindent\textbf{Open issues.} Usually, cellular networks are not energy-efficient as they consume energy in circuits, cooling systems, and also radiate in air. Hence, an energy-efficient deployment of a CRN in a cellular network is of utmost importance~\cite{DBLP:journals/cm/HongWWS14}. Further, there is a tradeoff between the spatial frequency reuse and the outage probability, which requires the selection of an efficient spectrum sensing algorithm~\cite{DBLP:journals/cm/ElSawyHK13}.

\subsection{Device-to-Device Communication Architectures}
\label{subsec:Device-to-Device Communication Architectures}
Device-to-Device (D2D) communication allows close proximity UEs to communicate with each other on a licensed cellular bandwidth without involving a MBS or with a very controlled involvement of a MBS. The standards and frameworks for D2D communication are in an early stage of research. In this section, we will review D2D communication networks in short. For a detailed review of D2D communication, interested readers may refer to~\cite{DBLP:journals/comsur/AsadiWM14}.

\newpage\noindent\textbf{Challenges in D2D communication.}
\begin{itemize}[noitemsep,nolistsep,leftmargin=.5cm]
  \item \textit{Interference management}: UEs involved in D2D communication, say D-UEs, face (or create) interference from (or to) other UEs, or from (or to) a BS, based on the selection of a DL or UL channel, respectively. The following types of interferences are investigated in~\cite{DBLP:journals/cm/WeiHQW14}:
      \begin{itemize}[noitemsep,nolistsep,leftmargin=.5cm]
      \item When using a DL channel: (\textit{i}) interference from BSs in the same cell, (\textit{ii}) interference from other co-channel D-UEs in the same cell, and (\textit{iii}) interference from BSs and co-channel D-UEs from other cells.
      \item When using a UL channel: (\textit{i}) interference from all co-channel C-UEs\footnote{{\scriptsize A UE that is not involved in D2D communication and communicates to a MBS, we call it a \textit{cellular user equipment} (C-UE) in this section.}} in the same cell and other cells, and (\textit{ii}) interference from all co-channel D-UEs in the same cell and other cells.
      \end{itemize}

\textit{Proposed solutions}: A simple solution may exist by implementing CRNs in D2D communication, as: D-UEs are considered as SUs and C-UEs are considered as PUs that should not be interfered. Consequently, any mechanism of CRNs can be implemented in D2D communication for interference removal.

  \item \textit{Resource allocation}: When UEs involved in D2D communication, it is required to allocate a sufficient amount of resources, particularly bandwidth and channels. However, the allocation of optimum resources to D-UEs must be carried out in a fashion that C-UEs do not have interference from D-UEs, and D-UEs can also communicate and exchange data efficiently~\cite{DBLP:journals/cm/WeiHQW14,DBLP:journals/cm/LiWHJC14}.

\textit{Proposed solutions}: SARA~\cite{DBLP:conf/globecom/ChenZZM14}, frame-by-frame and slot-by-slot channel allocation methods~\cite{DBLP:journals/wc/LiGMZ14}, and a social-aware channel allocation~\cite{DBLP:journals/cm/LiWHJC14}.

  \item \textit{Delay-sensitive processing}: Audio, video streaming, and online gaming, which are natural in close proximity UEs, require real-time and delay-sensitive processing. Hence, it is required to consider delay-sensitive and real-time processing in D2D communication~\cite{DBLP:journals/cm/WangL14}.

\textit{Proposed solutions}: Solutions based on channel state information (CSI) and QoS are provided in~\cite{DBLP:journals/cm/WangL14}.

  \item \textit{Pricing}: Sometimes a D-UE uses resources (\textit{e}.\textit{g}., battery and data storage) of other UEs for relaying information, where the other UE may charge for providing its resources. Hence, the design of a pricing model is needed, thereby a D-UE is not charged more money than that involved to communicate through a MBS.

      \textit{Proposed solutions}: Some solutions based on game theory, auction theory, and bargaining are suggested in~\cite{DBLP:journals/cm/TehraniUY14}.
\end{itemize}

\smallskip\noindent\textbf{D2D communication types.} D2D communication can be done in the following four ways~\cite{DBLP:journals/cm/TehraniUY14}, as follows:
\begin{enumerate}[noitemsep,nolistsep,leftmargin=.6cm]
  \item \textit{Device relaying with operator controlled link establishment} (DR-OC): A UE at the edge of a cell or in a poor coverage area can communicate with a MBS by relaying its information via other UEs, which are within the stronger coverage area and not at the edge; see Figure~\ref{fig:Two-tier architecture for 5G networks with small-cells}.

  \item \textit{Direct D2D communication with operator controlled link establishment} (DC-OC): Source and destination UEs communicate directly with each other without involving a MBS, but they are \textit{assisted} by the MBS for link establishment; see Figure~\ref{fig:Two-tier architecture for 5G networks with small-cells}.

  \item \textit{Device relaying with device controlled link establishment} (DR-DC): Source and destination UEs communicate \textit{through a relay} without involving a MBS, and they are also responsible for link establishment; see Figure~\ref{fig:Two-tier architecture for 5G networks with small-cells}.

  \item \textit{Direct D2D communication with device controlled link establishment} (DC-DC): Source and destination UEs communicate \textit{directly} with each other without involving a MBS, and they are also responsible for link establishment; see Figure~\ref{fig:Two-tier architecture for 5G networks with small-cells}.
\end{enumerate}
Note that DR-OC and DC-OC involve a MBS for resource allocation and call setup, and hence, prevent interference among devices to some extent.

\begin{wrapfigure}{r}{10cm}
\BBB\B
\begin{center}
\includegraphics[scale=0.38]{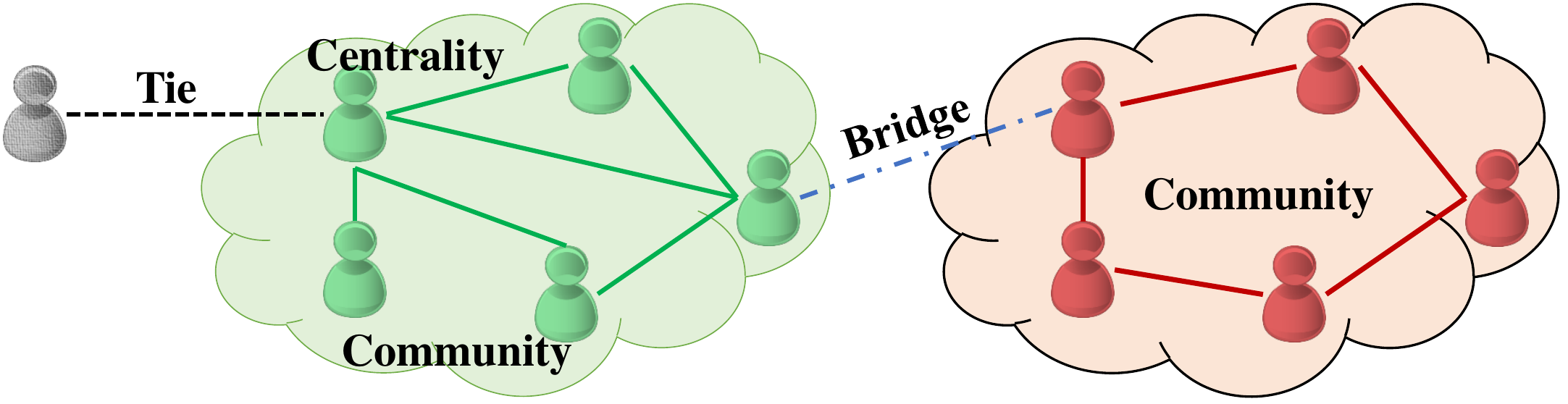}
\end{center}
\BB
\caption{A D2D communication architecture based on a social networking.}
\B
\label{fig:D2D communication architecture based on a social networking}
\end{wrapfigure}

Two types of coding schemes (or communication types) are described in~\cite{DBLP:journals/cm/WeiHQW14}: (\textit{i}) two-way relay channel (TRC), where a source and a destination communicate through a relay, and (\textit{ii}) multiple-access relay channel (MRC), where multiple sources communicate to a destination through a relay with direct links. Note that the workings of DR-OC and MRC, and DR-DC and TRC are identical. Two types of node discovery and D2D communication methods are also studied in~\cite{DBLP:journals/cm/LiWHJC14}, namely network-controlled approach and ad hoc network approach that work in a similar manner as DC-OC and DC-DC, respectively.

\smallskip\noindent\textbf{Resource allocation methods.} Now, we will review an architecture and some methods for resource allocation in D2D communication.

\textit{Social-Aware D2D Architecture}: As D2D communication is very efficient for close proximity UEs, keeping this fact in view, Li et al.~\cite{DBLP:journals/cm/LiWHJC14} suggested a social-aware D2D communication architecture based on the social networking. The architecture, see Figure~\ref{fig:D2D communication architecture based on a social networking}, has four major components as:

\begin{itemize}[noitemsep,nolistsep,leftmargin=.5cm]
  \item \textit{Ties}: They are similar to friend relations in a social media, and hence, may be used as a trust measurement between two UEs. Allocating more spectrum and energy resources to UEs with strong ties can increase the peer discovery ratio, avoid congestion, and improve spectral efficiency.

  \item \textit{Community}: It is similar to a group on Facebook and helps in allocating more resources to all the UEs in a community to decrease content duplication and increase the network throughput.

  \item \textit{Centrality}: It is similar to a node that has more communication links/friends in a social network. The concept of centrality in D2D communication reduces congestion and increases the network throughput by allocating more resources to a central node.

  \item \textit{Bridges}: They are similar to a connection between two communities. Hence, two devices forming a bridge can be allocated more resources as compared to other devices.
\end{itemize}

\textit{Channel Allocation Methods}: Two cooperative channel allocation methods, \textit{frame-by-frame} and \textit{slot-by-slot}, are given in~\cite{DBLP:journals/wc/LiGMZ14}. Consider three zones $A$, $B$, and $C$ with some UEs such that $A$ and $B$, $B$ and $C$ intersect, but $A$ and $C$ do not intersect, and $B$ holds a UE that communicates with UEs of $A$ and $C$. In the frame-by-frame channel allocation method, UEs of $A$ and $C$ do intra-zone communication at different frames, and the UEs of $B$ also communicate at different frames. However, in the slot-by-slot channel allocation method, UEs of $A$ and $C$ do intra-zone communication at an identical time, and of course, UEs of $B$ communicate at a different time. Both the methods improve the efficiency of frequency division multiplexing and increase the network throughput.

Hoang et al.~\cite{DBLP:conf/globecom/HoangLL14} provided an iterative algorithm for subcarrier\footnote{A subcarrier is a signal carried on a main radio signal, and hence, two signals are transmitted at an identical time.} and power allocation such that the minimal individual link data rates and proportionate fairness among D2D links are obtained. A 2-phase service-aware resource allocation scheme, called \textit{SARA}, is proposed~\cite{DBLP:conf/globecom/ChenZZM14}. In the first phase of SARA, resources are allocated on-demand to meet different service requirements of D-UEs, and in the second phase of SARA, the remaining resources are allocated to D-UEs such that the system throughput increases.

Wang et al.~\cite{DBLP:journals/cm/WangL14} provided a delay-aware and dynamic power control mechanism that adapts the transmit power of D-UEs based on instantaneous values of CSI, and hence, finds the urgency of the data flow. The dynamic power control selects a power control policy so that the long-term average delay and the long-term average power cost of all the flows minimize.

\smallskip\noindent\textbf{Advantages of D2D Communication.} D2D communication results in link reliability among D-UEs, a higher data rate to D-UEs, instant communication, an easy way for peer-to-peer file sharing, local voice services, local video streaming, local online gaming, an improved spectral efficiency, decreased power consumption of D-UEs, and the traffic offload from a MBS.

\smallskip\noindent\textbf{Open issues.}
\begin{itemize}[noitemsep,nolistsep,leftmargin=.5cm]
  \item \textit{Security and privacy}: In D2D communication, D-UEs may take helps from other UEs as relay nodes; hence, it is required to communicate and transfer data in secure and privacy-preserving manners. Consequently, the designing of energy-efficient and trust-making protocols is an open issue.

  \item \textit{Network coding scheme}: When D2D communication uses relay nodes, an efficient network coding scheme may be utilized for improving the throughput~\cite{DBLP:journals/cm/WeiHQW14}.

  \item \textit{Multi-mode selection}: In the current design of D2D communication, UEs can do either D2D communication or communication to a BS~\cite{DBLP:journals/cm/LiWHJC14}; however, it is not efficient. Hence, there is a need to design a system that allows a UE to engage the two types of communication modes (\textit{i}.\textit{e}., D2D communication and communication to a BS) simultaneously.
\end{itemize}

\subsection{Cloud-based Architectures}
\label{subsec:Cloud-based Architectures}
\begin{wrapfigure}{r}{7.5cm}
\BBB\BBB\BBB\B
\begin{center}
\includegraphics[scale=0.4]{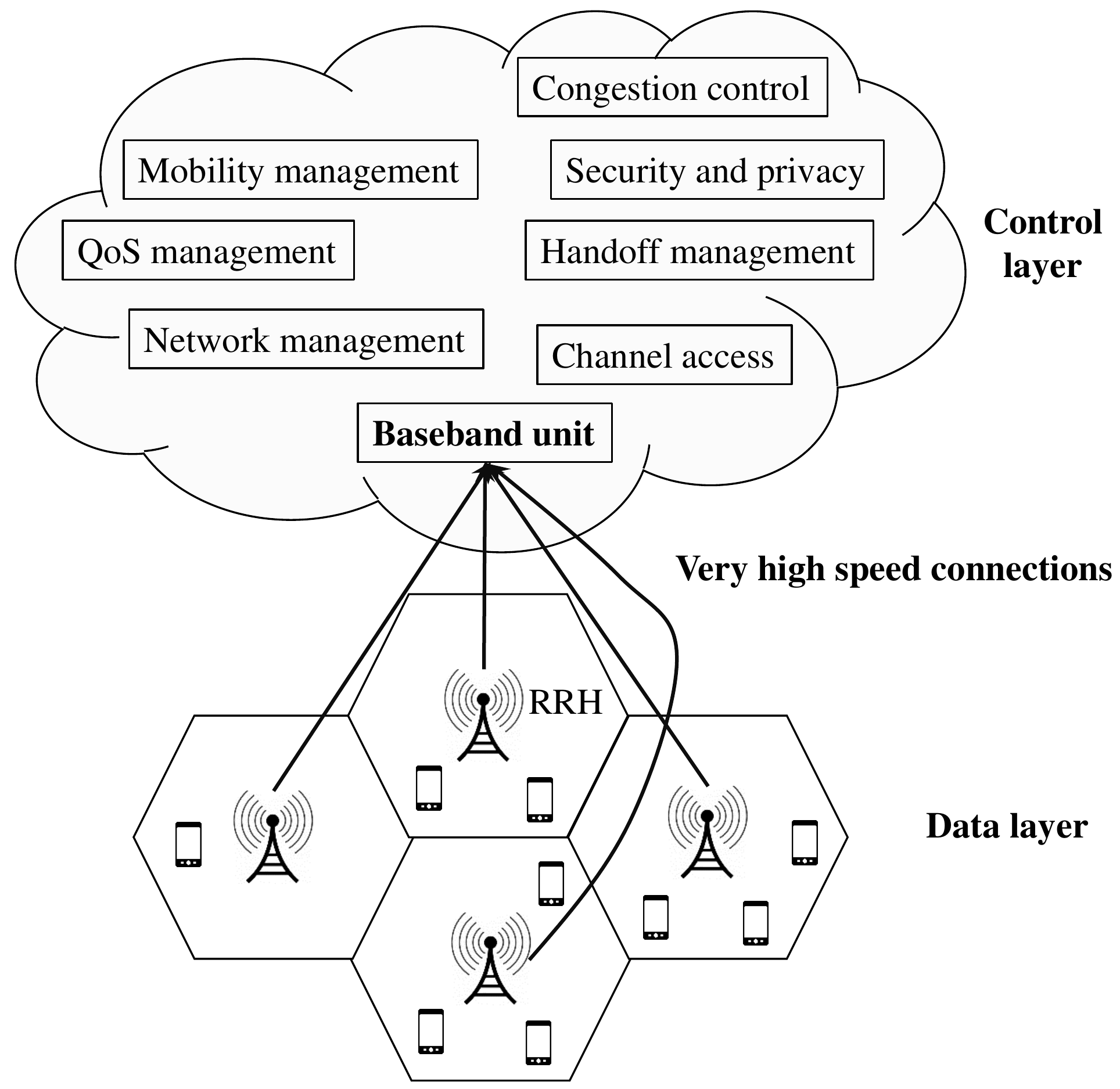}
\end{center}
\BBB
\caption{A basic cloud-based architecture for 5G networks.}
\BB\B
\label{fig:C-RAN-whole-cloud}
\BB
\end{wrapfigure}

Cloud computing~\cite{mell2011nist} infrastructure provides on-demand, easy, and scalable access to a shared pool of configurable resources, without worrying about the management of resources. The inclusion of the cloud in the mobile cellular communication can provide its benefits to the communication system. In this section, we will see cloud-based architectures or cloud-based radio access networks (C-RANs) for 5G networks. A detailed review of C-RANs is given in~\cite{DBLP:journals/comsur/CheckoCYSKBD15}.

\smallskip\noindent\textbf{The main idea of a C-RAN.} The first C-RAN is provided by China Mobile Research Institute~\cite{china}. The basic idea behind any C-RAN is to execute most of the functions of a MBS in the cloud, and hence, divide the functionality of a MBS into a \textit{control layer} and a \textit{data layer}; see Figure~\ref{fig:C-RAN-whole-cloud}. The functions of the control and the data layers are executed in a cloud and in a MBS, respectively. Thus, a C-RAN provides a \textit{dynamic} service allocation scheme for scaling the network without installing costly network devices.

Specifically, a MBS has two main components, as: (\textit{i}) a baseband unit (BBU, for implementing baseband processing using baseband processors), and (\textit{ii}) a remote radio head (RRH, for performing radio functions). In most of the C-RANs, BBUs are placed in the cloud and RRHs stay in MBSs. Thus, a C-RAN provides an easily scalable and flexible architecture. We will see advantages of C-RANs at the end of this section.

\smallskip\noindent\textbf{Challenges in the deployment of a C-RAN.}
\begin{itemize}[nolistsep,noitemsep,leftmargin=.5cm]
  \item \textit{An efficient fronthaul data transfer technique}: A flexible cloudification of the functions of a MBS comes at the cost of efficient fronthaul data transfer from RRHs to BBUs. The fast and efficient data transfer to the cloud has a proportionate impact on the performance of a C-RAN~\cite{DBLP:journals/cm/RostBDGLMSW14}.

  \item \textit{Real-time performance}: Since C-RANs will be used instead of a MBS that provides all the services to users, it is required to transfer and process all the data in the cloud as fast as a MBS can do; otherwise, it is hard to find solutions to real-time problems using a C-RAN~\cite{DBLP:journals/cm/RostBDGLMSW14}.

  \item \textit{Reliability}: The cloud provider does not ensure any guarantee of failure-free executions of their hardware and software. Thus, it is hard to simulate an error-free MBS using a C-RAN.

  \item \textit{Security}: The resources of the cloud are shared among several users and never be under the control of a single authority. Hence, a malicious user may easily access the control layer of a C-RAN, resulting in a more severe problem.

  \item \textit{Manageability}: It is clear that a non-secure C-RAN may be accessed by any cloud user, which poses an additional challenge in manageability of C-RANs. Further, the dynamic allocation of the cloud resources at a specific time interval is a critical issue; otherwise, a C-RAN may face additional latency~\cite{DBLP:journals/cm/HanGJL15}.

\end{itemize}

Now, we will see some C-RAN architectures in brief.

\smallskip\noindent\textbf{2-layered C-RAN architectures.} The authors~\cite{china} provided two C-RAN architectures based on the division of functionalities of a MBS, as: (\textit{i}) full centralized C-RAN, where a BBU and all the other higher level functionalities of a MBS are located in the cloud while a RRH is only located in the MBS, and (\textit{ii}) partially centralized C-RAN, where a RRH and some of the functionalities of a BBU are located in the MBS while all the remaining functions of the BBU and higher level functionalities of the MBS are located in the cloud. Thus, the authors~\cite{china} proposed the use of only two layers, namely a control layer and a data layer for implementing C-RANs, as follows:
\begin{enumerate}[nolistsep,noitemsep,leftmargin=.6cm]
  \item \textit{Data layer}: It contains heterogeneous physical resources (\textit{e}.\textit{g}., radio interface equipment) and performs signal processing tasks (\textit{e}.\textit{g}., channel decoding, demultiplexing, and fast Fourier transformation).
  \item \textit{Control layer}: It performs baseband processing and resource management (application delivery, QoS, real-time communication, seamless mobility, security, network management, regulation, and power control); see Figure~\ref{fig:C-RAN-whole-cloud}.
\end{enumerate}

Rost et al.~\cite{DBLP:journals/cm/RostBDGLMSW14} introduced RAN-as-a-service (RANaaS) concept, having the control and the data layers. However, in RANaaS, a cloud provides flexible and on-demand RAN functionalities (such as network management, congestion control, radio resource management, medium access control, and physical layer management), according to the network requirements and characteristics, unlike~\cite{china}. Hence, there is no need to split functionalities in advance to the control and the data layers, as a result RANaaS provides more elasticity.

Till now, it is clear that how a C-RAN will work. However, in order to achieve real-time performance, a RRH executing latency-critical applications may connect to a nearby small cloud while other RRHs that are not adhered to real-time applications may connect to a far larger cloud~\cite{DBLP:journals/cm/AgyapongISKB14}. SoftAir~\cite{ian2015} is also a two-layered C-RAN that performs mobility-management, resource-efficient network virtualization, and distributed and collaborative traffic balancing in the cloud.

\smallskip\noindent\textbf{3-layered C-RAN architectures.} The full-centralized C-RAN architecture~\cite{china} has some disadvantages, as: continuous exchange of raw baseband samples between the data and the control layers, and the control layer is usually far away from the data layer resulting in a processing delay.

In order to remove these disadvantages, Liu et al.~\cite{DBLP:journals/wc/LiuZZCN14} proposed \textit{convergence of cloud and cellular systems} (CONCERT). In this architecture, one more layer, called a \emph{software-defined service layer}, is introduced at the top of the control layer. The functioning of the layers in CONCERT is as follows:
\begin{enumerate}[nolistsep,noitemsep,leftmargin=.6cm]
  \item \textit{Data layer}: is identical to the full centralized C-RAN's data layer~\cite{china}, having RRHs with less powerful \textit{computational resources} for application level computations.
  \item \textit{Control layer}: works just as a logically centralized entity. The control layer coordinates with the data layer resources and presents them as virtual resources to the software-defined service layer. The control layer provides a few services as: radio interfacing management, wired networking management, and location-aware computing management to the data layer.
  \item \textit{Software-defined services layer}: works as a virtual BS and provides services (\textit{e}.\textit{g}., application delivery, QoS, real-time communication, seamless mobility, security, network management, regulation, and power control) to the data layer.
\end{enumerate}

Wu et al.~\cite{DBLP:journals/network/WuZHW15} enhanced C-RAN architecture~\cite{china} and RANaaS~\cite{DBLP:journals/cm/RostBDGLMSW14}, by moving the whole RAN to a cloud. The proposed architecture also has three layers, where the data layer and the control layer are same to the respective layers of C-RAN~\cite{china,DBLP:journals/wc/LiuZZCN14}. The third layer, called a \textit{service layer}, executes in the cloud and provides some more functionalities than the software-defined services layer of~\cite{DBLP:journals/cm/RostBDGLMSW14}, \textit{e}.\textit{g}., traffic management, the cell configuration, interference control, allocation of functional components to the physical elements, and video streaming services.

The authors~\cite{DBLP:journals/cm/YaziciKS14} proposed an all-software-defined network using three types of hierarchical network controllers, namely MBS controller, RAN controller, and network controller, where except the MBS controller all the others can be executed in the cloud, as follows:
\begin{enumerate}[nolistsep,noitemsep,leftmargin=.6cm]
  \item \textit{MBS controller}: usually stays nearby UEs, and performs wireless resource management and packet creation.
  \item \textit{RAN controller}: stays at the top of MBS controllers, and performs connectivity, RAT selection, handoff, QoS, policies, mobility management.
  \item \textit{Network controller}: stays at the top of RAN controllers, ensures end-to-end QoS, and establishes application-aware routes.
\end{enumerate}

\smallskip\noindent\textbf{Advantages of C-RANs in 5G networks.} C-RANs provide a variety of services as a software, power efficient, flexible, and scalable architecture for the future cellular communication. Here, we enlist some advantages of C-RANs, as follows:
\begin{itemize}[noitemsep,nolistsep,leftmargin=.5cm]
  \item \textit{An easy network management}: C-RANs facilitate on-demand installation of virtual resources and execute cloud-based resources that dynamically manage interference, traffic, load balance, mobility, and do coordinated signal processing~\cite{DBLP:journals/network/CaiYB14,DBLP:journals/cm/RostBDGLMSW14}.

  \item \textit{Reduce cost}: It is very costly and time-consuming to deploy and install a MBS to increase the network capacity. However, the deployment of C-RANs involves less cost, while it provides usual services like a MBS~\cite{DBLP:journals/network/WuZHW15}. As a result, operators are required to only deploy, install, and operate RRHs in MBSs.

  \item \textit{Save energy of UEs and a MBS}: C-RANs offload data-intensive computations from a MBS and may store data of UEs and MBSs. Consequently, C-RANs allow UEs and MBSs to offload their energy-consuming tasks to a nearby cloud, which saves energy of UEs and MBSs~\cite{DBLP:journals/spm/BarbarossaSL14}.

  \item \textit{Improved spectrum utilization}: A C-RAN enables sharing of CSI, traffic data, and control information of mobile services among participating MBSs, and hence, results in increased cooperation among MBSs and reduced interference~\cite{DBLP:journals/network/WuZHW15}.
\end{itemize}

\smallskip\noindent\textbf{Open issues.} Transferring data from RRHs to BBUs, \textit{i}.\textit{e}., from the data layer to the control layer, is a crucial step based on the selection of the functions of a MBS that has to be sent to a cloud, resulting in the minimal data movement in the network~\cite{DBLP:journals/cm/RostBDGLMSW14}. However, the selection of functions to be executed in a cloud and a MBS is a non-trivial affair~\cite{DBLP:journals/cm/RostBDGLMSW14}. The security and privacy issues involved in the cloud computing effect C-RANs, and hence, the development of a C-RAN has to deal with inherent challenges associated with the cloud and the wireless cellular communication simultaneously.

\subsection{Energy-Efficient Architectures for 5G Networks}
\label{subsec:Energy-Efficient Architectures for 5G Networks}
Energy-efficient infrastructures are a vital goal of 5G networks. Researchers have proposed a few ways of reducing energy in the infrastructure. Rowell et al.~\cite{DBLP:journals/cm/IRHXLP14} considered a joint optimization of energy-efficiency and spectral-efficiency. A user-centric 5G network is suggested in~\cite{DBLP:journals/cm/IRHXLP14} so that UEs are allowed to select UL and DL channels from different BSs depending on the load, channel conditions, services and application requirements. In a similar manner, decoupling of signaling and data is useful for energy saving; for example, a MBS may become a signaling BS while SBSs may serve all data requests. Thus, when there is no data traffic in a SBS, it can be turned off. A similar approach for decoupling of signaling and data is presented in~\cite{DBLP:journals/cm/ZhangZWZIPLC15}. However, in~\cite{DBLP:journals/cm/ZhangZWZIPLC15}, a UE gets connected to a SBS according to instructions by a MBS, and hence, it results in less energy consumption at UEs' side due to less interference, faster small-cells' discovery, and MBS-assisted handover.

Hu and Qian~\cite{DBLP:journals/cm/HuQ14} provided an energy-efficient C-RAN architecture in a manner that RRHs serve almost a same number of UEs. They also present an interference management approach so that the power consumption of SBSs and MBSs can be decreased. Like Rowell et al.~\cite{DBLP:journals/cm/IRHXLP14}, Hu and Qian~\cite{DBLP:journals/cm/HuQ14} also suggested that the association of a UE cannot be done based on entirely a DL channel or a UL channel, and a UE must consider both the channels at the time of association with a BS. Lin et al.~\cite{DBLP:journals/network/LiuZYX15} suggested to include an energy harvesting device (to collect energy) and a spectrum harvesting controller (to collect spectrum) at SBSs.

\section{Implementation Issues in 5G Networks}
\label{sec:Management Issues in 5G Networks}
In this section, we will see issues regarding the interference, handoff, QoS, load balancing, and channel access management in the context of 5G networks.

\subsection{Interference Management in 5G Networks}
\label{subsec:Interference Management in 5G Networks}
We have already seen challenges in interference management (Section~\ref{section:Challenges in the Development of 5G}). In this section, we will review some techniques/methods for interference management in 5G networks.

Nam et al.~\cite{DBLP:journals/cm/NamBLK14} handled UE-side interference by using a new type of receiver equipment, called an advanced receiver, which detects, decodes, and removes interference from receiving signals. In addition, the network-side interference is managed by a joint scheduling, which selects each UE according to the resources needed (\textit{e}.\textit{g}., time, frequency, transmission rate, and schemes of multiple cells) for its association with a BS. Hence, the joint scheduling, which can be implemented in a centralized or distributed manner, requires a coordination mechanism among the neighboring cells.

Hossain et al.~\cite{DBLP:journals/wc/HossainRTA14} proposed distributed cell access and power control (CAPC) schemes for handling interference in multi-tier architectures. CAPC involves: (\textit{i}) prioritized power control (PPC), which assumes that UEs working under a SBS have a low-priority than UEs working under a MBS, and hence, low-priority UEs set their power so that the resulting interference must not exceed a predefined threshold; (\textit{ii}) cell association (CA), which regards dynamic values of resources, traffic, distance to a MBS, and available channels at a MBS for selecting a MBS with the optimum values of the parameters; and (\textit{iii}) resource-aware CA and PPC (RCA-PPC), which is a combination of the first two approaches and allows a UE to connect simultaneously with multiple BSs for a UL channel and a DL channel based on criteria of PPC and CA.

Hong et al.~\cite{DBLP:journals/cm/HongBCJMKL14} suggested to use self-interference cancellation (SIC) in small-cells' networks. As we have seen that SBSs require methods to transfer backhaul data to a MBS (Section~\ref{subsubsec:Backhaul data transfer from small-cells}), the use of SIC can eliminate the need of such methods and result in \textit{self-backhauled small-cells} (SSCs). SSCs use SIC for providing services and backhaul data transfer, and more importantly, they gain almost the same performance as having a small-cell connected with a wired optical fiber. It works as: in the DL channel, a SBS may receive from a MBS and simultaneously transmit to UEs. In the UL channel, a SBS may receive from UEs and simultaneously transmit data to the MBS. Therefore, a small-cell can completely remove the need of a separate backhaul data transfer method, resulting in reduced cost.

The authors~\cite{DBLP:journals/cm/IRHXLP14} suggested that the measurement of inter-user interference channel, and then, the allocation of UL and DL channels by a MBS can mitigate \textit{inter-user UL-to-DL interference} in a \textit{single-cell} full duplex radio network. However, in the case of a \textit{multi-cell} full duplex radio network, interference mitigation becomes more complex, because of the existence of interference in UL and DL channels between multi-cells' UEs that work on identical frequency and time.

\smallskip\noindent\textbf{Open issues.} Interference cancellation in a full duplex radio network is still an open problem for multi-cell. The design of algorithms for inter-BS DL-to-UL-interference and inter-user UL-to-DL-interference cancellation in a full duplex radio network is still open and to be explored.

\subsection{Handoff Management in 5G Networks}
\label{subsec:Handoff Management in 5G Networks}
Handoff provides a way to UEs connected to a BS to move to another BS without disconnecting their sessions.

\smallskip\noindent\textbf{Challenges in the handoff process in 5G networks.} The handoff management in 5G networks has inherent challenges associated with the current cellular networks, \textit{e}.\textit{g}., minimum latency, improved routing, security, and less uncertainty of having no services. The network densification, very high mobility, the zero latency, and accessing multi-RATs make handoff management in 5G networks harder. Also, the current cellular networks do not provide efficient load balancing for a BS at the time of handoff. For example, movement of UEs from houses to offices in the morning creates a load imbalance at BSs of respective areas.

\smallskip\noindent\textbf{Types of handoff in 5G networks.} Three types of handoffs are presented in the context of 5G networks, as follows:
\begin{enumerate}[noitemsep,nolistsep,leftmargin=.6cm]
  \item \textit{Intra-macrocell handoff}: refers to handoff between small-cells that are working under a single MBS~\cite{DBLP:journals/tvt/SongFY14}.
  \item \textit{Inter-macrocell handoff}: refers to handoff between macrocells. It may also lead to handoff between two small-cells that are working under different MBSs. Note that if the handover between small-cells of two different MBSs is not done properly, then the inter-macrocell handoff also fails~\cite{DBLP:journals/tvt/SongFY14}.
  \item \textit{Multi-RATs handoff}: refers to handoff of a UE from a RAT to other RAT.
\end{enumerate}

Song et al.~\cite{DBLP:journals/tvt/SongFY14} provided a handoff mechanism for highly mobile users, where a UE sends some parameters (\textit{e}.\textit{g}. QoS, signal-to-interference ratio (SIR), and time to handoff) in a measurement report to the current MBS. SIR is considered as a primary factor for finding a situation for an initiation of the handoff. The Gray system model predicts the $(N + 1)^{th}$ measurement report from the $N^{th}$ measurement report. The predicted value is used for the final decision for the handover process. Zhang et al.~\cite{DBLP:journals/cm/ZhangZWZIPLC15} proposed a handoff mechanism assisted by a MBS. The MBS collects several parameters from UEs, and if the MBS finds the values of the received parameters below a threshold, then it finds a new SBS or MBS for handoff and informs to the UEs.

For handoff over different RATs, Orsino et al.~\cite{DBLP:journals/corr/OrsinoAMI15} proposed a handoff procedure so that a UE can select the most suitable RAT without any performance loss. A UE collects received signal strength (\textit{i}.\textit{e}., RSRP) or quality (\textit{i}.\textit{e}., RSRQ) from the current MBS, and then, it initiates handoff if RSRQ is below a threshold. The UE collects several parameters (\textit{e}.\textit{g}., transmitted power, the cell's traffic load, and UE requested spectral efficiency) from adjacent BSs, and then, selects the most suitable BS.

Duan et al.~\cite{DBLP:journals/cm/DuanW15} provided an authenticated handoff procedure for C-RANs and multi-tier 5G networks. The control layer holds an authentication handover module (AHM) for monitoring and predicting the future location of UEs (based on the current location) and preparing relevant cells before UEs' arrival in that. The AHM holds a master public-private key pair with each RRH, which are authenticated by the AHM in off-peak hours, and UEs are verified before accessing the network services by RRHs. UEs sends ID, the physical layer's attributes, location, speed, and direction to the control layer in a secure manner for the handoff process. The proposed approach reduces the risk of impersonation and man-in-the-middle attacks. Giust et al.~\cite{DBLP:journals/cm/GiustCB15} provided three distributed mobility management protocols, the first is based on the existing Proxy Mobile IPv6 (PMIPv6), the second is based on SDN, and the third is based on routing protocols.

\smallskip\noindent\textbf{Open issues.} Handoff mechanisms are yet to be explored. It will be interesting to find solutions to extremely dense HetNets. Furthermore, the handoff process may also create interference to other UEs; hence, it is required to develop algorithms while considering different types of interferences in 5G networks and a tradeoff between the number of handoffs and the level of interference in the network~\cite{DBLP:journals/twc/LeuMT08}.

\subsection{QoS Management in 5G Networks}
\label{subsec:QoS Management in 5G Networks}
We have already seen challenges in QoS management in Section~\ref{section:Challenges in the Development of 5G}. In this section, we will review some techniques/methods for QoS management in 5G networks.

Zhang et al.~\cite{DBLP:journals/network/ZhangCZ14} provided a mechanism for different delay-bounded QoS for various types of services, applications, and users having diverse requirements, called heterogeneous statistical delay-bounded QoS provisioning (HSP). HSP maximizes the aggregate effective capacity of different types of 5G architectures, reviewed in Sections~\ref{subsec:Two-tier architectures},~\ref{subsec:Cognitive Radio Network based Architectures}, and~\ref{subsec:Device-to-Device Communication Architectures}. HSP~\cite{DBLP:journals/network/ZhangCZ14} algorithm claims better performance over other approaches; however, it imposes new challenges in terms of the assignment of different resources for different links under the cover of delay-bounded QoS requirements.

The authors~\cite{DBLP:journals/network/ZhouZLZCNZ14} proposed the deployment of a quality management element (QME) in the cloud for monitoring inter-UEs and inter-layer (the control and the data layers in C-RANs) QoS. RRUs send wireless information (\textit{e}.\textit{g}., CSI, reference signal received quality, and resource block utilization) to the QME. Consequently, the QME executes service control algorithms (to manage QoS and some other activities like traffic offloading and customized scheduling), and then, sends scheduling strategies to the RRUs to achieve a desired level of QoS.

Kim et al.~\cite{DBLP:conf/icc/KimM14} provided a routing algorithm for multi-hop D2D communication. The algorithm takes into account different QoS for each link so that it can achieve better performance than max-min routing algorithms. The algorithm increases flow until it provides the desired QoS or reaches the maximum capacity of the link. Since the algorithm considers individual links, there is a high probability that some of the channels will serve multiple-links with the desired QoS. Zhou et al.~\cite{DBLP:conf/wcsp/ZhouLWYG14} provided a QoS-aware and energy-efficient resource allocation algorithm for DL channels, where UEs are allocated an identical power in one case and non-identical power in the second case. The algorithm maximizes energy-efficiency while minimizes transmit power. Hu and Qian~\cite{DBLP:journals/cm/HuQ14} suggested that a UE must consider the source data rate, delay bound, and delay-violation probability before connecting to a MBS.

\smallskip\noindent\textbf{Open issues.} The 5G networks are supposed to satisfy the highest level of QoS. The tactile Internet~\cite{ti2014} requires the best QoS, especially, latency of the order of 1 millisecond for senses such as touching, seeing, and hearing objects far away, as precise as human perception. However, the current proposed architectures do not support efficient tactile Internet services. In future, it would be a promising area as to encode senses, exchange data satisfying the zero latency, and enable the user to receive the sensation.

\subsection{Load balancing in 5G Networks}
\label{subsec:Load balancing in 5G Networks}
Load balancing means allocation of resources to a cell such that all the users meet their demands. It is an important issue in the cellular wireless networks. In a 2-tier architecture, discussed in Section~\ref{subsec:Two-tier architectures}, user offloading to a small-cell is useless if there is no resource partitioning~\cite{DBLP:journals/twc/SinghA14}. Singh and Andrews~\cite{DBLP:journals/twc/SinghA14} provided an analytical and tractable framework for modeling and analyzing joint resource partitioning and offloading in a 2-tier architecture. Hossain et al.~\cite{DBLP:journals/wc/HossainRTA14} provided a technique for cell association based on dynamic resources and traffic in a cell, as we have already discussed in Section~\ref{subsec:Interference Management in 5G Networks}. In a fast moving vehicle, \textit{e}.\textit{g}., a train, it is very hard to allocate resources without any service interruption at all. A distributed load balancing algorithm for fast moving vehicles is presented in~\cite{DBLP:conf/5gu/GorattiSPS14}. Load balancing architectures for D2D communication are given in~\cite{DBLP:journals/cm/LiWHJC14,DBLP:conf/globecom/ChenZZM14,DBLP:conf/globecom/HoangLL14}, which we discussed in Section~\ref{subsec:Device-to-Device Communication Architectures}. Interested readers may refer to~\cite{DBLP:journals/wc/AndrewsSYLD14} for further details of load balancing in 5G networks.

\subsection{Channel Access Control Management in 5G Network}
\label{subsec:Medium Access Control Management for 5G Network}
Channel access protocols allow several UEs to share a transmission channel without any collision while utilizing the maximum channel capacity.

\smallskip\noindent\textbf{Challenges in channel access control management in 5G networks.} Channel access control management in 5G networks faces inherent challenges associated with the current cellular networks, \textit{e}.\textit{g}., synchronization, fairness, adaptive rate control, resource reservation, real-time traffic support, scalability, throughput, and delay. In addition, providing the currently available best channel in 5G networks is prone to additional challenges, as: high mobility of UEs, working at the higher frequencies ($>$ 3 GHz), different RATs, dense networks, high QoS, high link reliability, and the zero latency for applications and services.

\smallskip The authors~\cite{rappport15} proposed a frame-based medium access control (FD-MAC) protocol for mmWave-based small-cells. FD-MAC consists of two phases, as: (\textit{i}) scheduling phase, when a SBS collects the traffic demands from the supported UEs and computes a schedule for data transmission, and (\textit{ii}) transmission phase, when the UEs start concurrent transmissions following the schedule. A schedule, which is computed using a graph-edge coloring algorithm, consists of a sequence of topologies and a sequence of time intervals for indicating how long each topology should sustain.

Liu et al.~\cite{DBLP:journals/network/LiuZYX15} provided MAC protocols for UEs of small-cells and macrocells. Two types of MAC protocols for SBS's UEs are suggested, as: contention-based random channel access (CRCA), where UEs randomly access the channel and send messages if the channel is available, and reservation-based channel access (RCA), which uses time division multiple access. For macrocells' UEs, they provided a MAC protocol for CRN-based 5G networks, where SUs sense a licensed channel until it is free or the residual energy of SUs exceeds a predetermined threshold, for saving their battery. Further, they evaluated a tradeoff between the network throughput and sensing overhead.

The authors~\cite{DBLP:journals/cm/ElSawyHK13} suggested two CRN-based channel access techniques for cognitive SBSs. The first channel access scheme is termed as contention-resolution-based channel access (CCA), which is similar to CRCA~\cite{DBLP:journals/network/LiuZYX15}, is based on carrier sense multiple access. The other channel access scheme is termed as uncoordinated aggressive channel access (UCA), which works identically as RCA~\cite{DBLP:journals/network/LiuZYX15}, for aggressively using channels in a small-cell for increasing opportunistic spectrum access performance. Nikopour et al.~\cite{DBLP:conf/globecom/NikopourYBAHBM14} provided a multi-user sparse code multiple access (MU-SCMA) for increasing DL's spectral efficiency. MU-SCMA does not require the complete CSI, and hence, provides high data rate and the robustness to mobility. Nikopour et al.~\cite{DBLP:conf/globecom/NikopourYBAHBM14} also provided an uplink contention based SCMA for massive connectivity, data transmission with low signaling overhead, low delay, and diverse traffic connectivity requirements.

\smallskip\noindent\textbf{Open issues.} The current channel access protocols do not regard QoS and latency challenges in 5G networks. Hence, there is a scope of designing algorithms for finding multiple reliable links with the desired QoS and the zero latency.

\subsection{Security and Privacy Management in 5G Networks}
\label{subsec:Security and Privacy Management in 5G Networks}
In this section, we present security and privacy related challenges and a discussion of security and privacy protocols in the context of 5G networks.

\smallskip\noindent\textbf{Challenges in security and privacy in 5G networks.} \textit{Authentication} is a vital issue in any network. Due to the zero latency guarantee of 5G networks, authentication of UEs and network devices is very challenging, since the current authentication mechanisms use an authentication server that takes hundreds of milliseconds delay in a preliminary authentication phase~\cite{DBLP:journals/cm/DuanW15}. A \textit{fast and frequent handover} of UEs over small-cells requires for a robust, efficient, and secure handoff process for transferring context information~\cite{DBLP:journals/cm/DuanW15}. Security to \textit{multi-RATs} selection is also challenging, since each RAT has its own challenges and certain methods to provide security; clearly, there is a need to provide overlapped security solutions across the various types of RATs. \textit{C-RANs} also inherit all the challenges associated with the cloud computing and wireless networks. In addition, several other challenges (\textit{e}.\textit{g}., authorization and access control of UEs, availability of the network, confidentiality of communication and data transfer, integrity of communication and data transmission, accounting and auditing of a task, low computation complexity, and communication cost) require sophisticated solutions to make a secure 5G network. The security and latency are correlated as a higher level of security and privacy results in increased latencies. Therefore, the communication satisfying the zero latency is cumbersome when combined with secure and privacy-preserving 5G networks.

\smallskip Monitoring is suggested for securing the network and detecting intruders~\cite{DBLP:journals/wpc/Ulltveit-MoeOK11,gai2015intrusion}. However, monitoring a large number of UEs (by a trusted authority) is not a trivial task; hence, we do not see monitoring as a preferred way to secure networks. Yang et al.~\cite{DBLP:journals/cm/YangWGEYR15} focused on the physical layer security, which is independent of computational complexity and easily handles any number of devices. The physical layer security protocol considers locations of UEs and provides the best way for UEs for securely selecting a MBS or a SBS without overloading the network. Tehrani et al.~\cite{DBLP:journals/cm/TehraniUY14} provided a method for secure and private D2D communications, called close access, where D-UEs have a list of other trusted D-UEs devices, and all such UEs can communicate directly using an encryption scheme while the remaining UEs not in the list utilize a MBS-assisted communication. An encryption based video sharing scheme is also presented in~\cite{DBLP:journals/network/LiuASLLS15}. Kantola et al.~\cite{policy-based} proposed a policy-based scheme that can prevent DoS and spoofing attacks. A solution to secure handoff is given in~\cite{DBLP:journals/cm/DuanW15} and discussed in Section~\ref{subsec:Handoff Management in 5G Networks}.

\smallskip\noindent\textbf{Open issues.} The current security and privacy solutions to 5G networks are not impressive and, hopefully, unable to handle massive connections. We can clearly visualize a potential scope for developing \textit{latency-aware protocols} along with security awareness that must consider secure data transmission, end-to-end security, secure and private storage, threats resistant UEs, and valid network and software access.

\section{Methodologies and Technologies for 5G Networks}
\label{sec:Methodology and Technology for 5G Networks}
It is already mentioned in Section~\ref{section:introduction} that the development of 5G networks requires the design and implementation of new methodologies, techniques, and architectures. We have reviewed some of the methodologies and technologies in the previous sections, such as: (\textit{i}) full duplex radios in Sections~\ref{section:Challenges in the Development of 5G} and~\ref{subsec:Interference Management in 5G Networks}, (\textit{ii}) CRNs in Section~\ref{subsec:Cognitive Radio Network based Architectures}, (\textit{iii}) D2D communication in Section~\ref{subsec:Device-to-Device Communication Architectures}, (\textit{iv}) multi-tier heterogeneous deployment or dense-deployment techniques in Sections~\ref{subsec:Two-tier architectures},~\ref{subsec:Cognitive Radio Network based Architectures}, and~\ref{subsec:Device-to-Device Communication Architectures}, (\textit{v}) C-RANs in Section~\ref{subsec:Cloud-based Architectures}, (\textit{vi}) `green' communication systems in Section~\ref{subsec:Energy-Efficient Architectures for 5G Networks}, and (\textit{vii}) techniques related to interference, QoS, handoff, channel access, load balancing in Section~\ref{sec:Management Issues in 5G Networks}.

In this section, we will briefly describe some techniques that are mentioned but not explained in earlier sections.

\smallskip\noindent\textbf{Self-Interference Cancellation (SIC).}
When a full duplex radio receives signals from another radio, it also receives interference signals by its own transmission, resulting in self-interference. Hence, a full duplex radio has to implement techniques to cancel self-interference~\cite{DBLP:conf/sigcomm/BharadiaMK13,DBLP:conf/globecom/HanIXPP14,DBLP:journals/cm/ZhangCLVH15}. SIC techniques are classified into passive and active cancellations; see~\cite{DBLP:journals/cm/ZhangCLVH15}. As advantages, the implementation of SIC enables seamless global roaming, high-throughput services, and low-latency applications in a cost effective manner~\cite{DBLP:journals/cm/HongBCJMKL14}.

\smallskip\noindent\textbf{Downlink and Uplink Decoupling (DUD).} In the current cellular networks, a UE is associated with a BS based on the received signal power in its DL channel, and then, uses the same BS for UL channel transmission~\cite{DBLP:journals/wcl/SmiljkovikjPG15}. DUD allows a UE to select a DL channel and a UL channel from two different BSs, based on the link quality, the cell load, and the cell backhaul capacity~\cite{DBLP:conf/globecom/ElshaerBDI14,DBLP:journals/corr/ElshaerBDI14a}. Therefore, a UE may have the DL channel connected through a BS and the UL channel connected through a different BS, resulting in a user-centric 5G architecture~\cite{DBLP:journals/cm/IRHXLP14} and improving the capacity of UL channels, which is a prime concern.

\smallskip\noindent\textbf{Network Function Virtualization (NFV).} NFV~\cite{DBLP:journals/cm/HanGJL15} implements network functions such as network address translation, firewalls, intrusion detection, domain name service, the traffic load management, and caching through software running on commodity servers. However, the conventional networks implement these functions on dedicated and application specific servers. Hence, NFV decreases the burden on network operators by not updating dedicated servers/hardware, thereby saves cost.

\smallskip\noindent\textbf{Software-Defined Networking (SDN).} SDN architectures~\cite{DBLP:journals/cm/KimF13,DBLP:journals/cn/FarhadiLN15} partition network control functions and data forwarding functions, thereby the network control functions are programmable, and the network infrastructure handles applications and network services. SDN architectures can be divided into three parts, as: (\textit{i}) \textit{the software controller}: holds network control functions such as the network manager, APIs, network operating system, and maintaining the global view of the network; (\textit{ii}) \textit{the southbound part}: provides an interface and a protocol between the controller and SDN-enable infrastructure, where OpenFlow~\cite{DBLP:journals/ccr/McKeownABPPRST08} is the most famous protocol that provides communication between the controller and the southbound part; (\textit{iii}) \textit{the northbound part}: provides an interface between SDN applications and the controller~\cite{DBLP:journals/network/Hernandez-Valencia15}. Interested readers may refer to~\cite{DBLP:journals/network/CaoHLWL15,DBLP:journals/comcom/XieGHQL15,DBLP:journals/pieee/KreutzRVRAU15,DBLP:journals/cn/FarhadiLN15,DBLP:journals/cm/ArslanSR15,Alsmadi201579} to find details of SDN, challenges in SDN, and applications of SDN.

Note that SDN, NFV, and C-RANs offload functionalities to software running on commodity servers. However, SDN separates network control functions from data forwarding functions, while NFV implements network functions in software~\cite{DBLP:journals/access/WangHY14}. Besides that C-RANs integrate both SDN and NFV to meet the scalability and flexibility requirements in the future mobile networks~\cite{DBLP:journals/access/WangHY14}.

\smallskip\noindent\textbf{Millimeter Waves (mmWave).} The current wireless bandwidth is not able to support a huge number of UEs in 5G networks. Hence, researchers are looking at 30 GHz to 300 GHz frequency bands, where mmWave communication is proposed for achieving high-speed data transfer. The current research focuses on 28 GHz band, 38 GHz band, 60 GHz band, and the E-band (71–76 GHz and 81–86 GHz)~\cite{DBLP:journals/corr/NiuLJSV15}. However, mmWave has several challenges at the physical, MAC, and network layers. There are a number of papers about mmWave, and hence, we are not discussing mmWave in details. Interested readers may refer to~\cite{DBLP:journals/access/RappaportSMZAWWSSG13,DBLP:journals/cm/BoccardiHLMP14,DBLP:journals/corr/NiuLJSV15}.

\smallskip\noindent\textbf{Machine-to-Machine (M2M) communication.} M2M communication refers to the communication between (network) devices (\textit{e}.\textit{g}., sensors/monitoring-devices with a cloud) without human intervention. Some examples of M2M communication are intelligent transport systems, health measurement, monitoring of buildings, gas and oil pipelines, intelligent retail systems, and security and safety systems. However, the development of M2M communication involves several challenges to be handled in the future, as follows: connectivity of massive devices, bursty data, the zero latency, scalability in terms of supporting devices, technologies and diverse applications, fast and reliable delivery of messages, and cost of devices. In addition, the development of efficient algorithms for location, time, group, priority, and multi-hop data transmission management for M2M communication is needed~\cite{outlook7,DBLP:journals/cm/StenumgaardCFA13}. Interested readers may refer to~\cite{anton2014machine,DBLP:journals/wc/FanHK14,DBLP:journals/wcl/MaduenoSP14,DBLP:journals/cm/CondoluciDAMZ15,DBLP:conf/icin/RatasukPLGU15}.

\smallskip\noindent\textbf{Massive MIMO (mMIMO).} mMIMO systems are also known as large-scale antenna systems, very large MIMO, hyper-MIMO, and full-dimension MIMO~\cite{DBLP:journals/cm/LarssonETM14,DBLP:journals/cm/NamNSL0KL13}. mMIMO systems use antenna arrays with hundreds of antennas at MBS for simultaneously serving many UEs with a single antenna in identical time and frequency. Hence, expensive equipment are mounted on a MBS~\cite{DBLP:journals/jstsp/LuLSAZ14}. A mMIMO system relies on spatial multiplexing, which in turn relies on the channel knowledge at MBS, on both UL and DL channels. A mMIMO system reduces latency and energy, simplifies the MAC layer, shows robustness against intentional jamming, and increases the capacity due to spatial multiplexing.

\smallskip\noindent\textbf{Visual Light Communication (VLC).} VLC is a high-speed data transfer medium for short range LOS optical links in the future cellular networks~\cite{DBLP:journals/network/WuW14}. The light-emitting diodes (LEDs) provide VLC using amplitude modulation at higher frequencies and achieve higher data rates while keeping the LED’s primary illumination function unaffected. VLC can be used for outdoor applications, where high power laser-based equipment provide transmission links, and for indoor application, where LEDs provide short distance transmission links. VLC is an energy-efficient technology, works on a wider range of unregulated frequency bands, shows high spatial reuse, and inherits security due to LOS. However, VLC is sensitive to sunlight and not able to work for a long range without LOS, and hence, confined coverage~\cite{DBLP:journals/cm/GrobePHSLHKJL13,DBLP:journals/cm/JovicicLR13}. The implementation of VLC has still some unanswered questions, such as: how VLC will work in a long range without LOS and will it work for backhaul data transfer in multihop?

\smallskip\noindent\textbf{Fast caching.} Caching is a way for storing temporary data for reducing data access from slow memory or the network. In a network, content caching\footnote{{\scriptsize \url{https://developer.akamai.com/stuff/Caching/Content_Caching.html}}} is popular and answers the request while responding in place of the application servers as a proxy, thereby reducing the amount of hits that are directly sent to the ultimate backend server. In caching, three decisions are prominent as: what to cache, where to cache, and how to cache? The authors~\cite{DBLP:journals/cm/WangCTKL14,DBLP:journals/cm/BoccardiHLMP14} suggested that UEs will have enough memory in the future, and they can work as a cache for any other UE, since a small amount of popular data requires to be cached. Wang et al.~\cite{DBLP:journals/cm/WangCTKL14} provided a caching mechanism based content centric networking (CCN), assuming 5G networks will include CCN-capable gateways, routers, and MBSs. CCN provides in-network data storage, also known as \textit{universal caching}, at every node in the network. The cached data is uniquely identified at each node. Accordingly, a user can request for a particular content from the content cache of any device within the network, or the request is forwarded to the actual source of content storage.

Interested readers may refer to~\cite{6730679,DBLP:journals/cm/BoccardiHLMP14} for finding details about some of the above mentioned methodologies and technologies.

\section{Applications of 5G Networks}
\label{sec:Applications of 5G Networks}
The zero latency, high speed data transfer, and ubiquitous connectivity are the salient features of 5G networks that are expected to serve a wide range of applications and services. In this section, we enumerate the most prominent applications of 5G networks, as follows:

%\begin{itemize}[nolistsep,noitemsep]
\smallskip\noindent\textbf{Personal usages.} This domain of 5G networks would be capable of supporting a wide range of UEs, from scalable to heterogeneous devices. Also the data demands (\textit{e}.\textit{g}., multimedia data, voice communication, and Web surfing), would be satisfied while keeping the QoS requirements.

\smallskip\noindent\textbf{Virtualized homes.} Due to C-RAN architectures, users may have only low cost UEs (\textit{e}.\textit{g}., set-top box for TVs and residential gateways for accessing the Internet) with services of the physical and data link layers. All the other higher layers' applications may move to the cloud for universal access and outsourced computation services~\cite{DBLP:journals/cm/HanGJL15}.

\smallskip\noindent\textbf{Smart societies.} It is an abstract term for connected virtualized homes, offices, and stores. Accordingly, every digital and electronic services/appliances, \textit{e}.\textit{g}., temperature maintenance, warning alarms, printers, LCDs, air conditioners, physical workout equipment, and door locks, would be interconnected in a way that the collaborative actions would enhance the user experience. Similarly, smart stores would assist in filtering out irrelevant product details, sale advertisements, and item suggestions on the go.

\smallskip\noindent\textbf{Smart grids.} The smart grids would decentralize the energy distribution and better analyze the energy consumption. This would allow the smart grids to improve efficiency and economic benefits. The 5G networks would allow a rapid and frequent statistical data observation, analysis, and fetching from remote sensors and would adjust the energy distribution accordingly.

\smallskip\noindent\textbf{The tactile Internet.} The tactile Internet improves the user experience in a virtual environment to an extent of only milliseconds of interaction latency. The futuristic applications such as automated vehicle platooning, self-organizing transportation, the ability to acquire a virtual sense for physically challenged patients, synchronized remote smart-grids, remote robotics, and image processing with customized/panoramic view would use the tactile Internet protocols~\cite{ti2014}.

\smallskip\noindent\textbf{Automation.} Self-driving vehicles would take place in the near future, and as a requirement, vehicles would communicate with each other in real-time. Moreover, they would communicate with other devices on the roads, homes, and offices with a requirement of almost zero latency. Hence, an interconnected vehicular environment would provide a safe and efficient integration with other information systems.

\smallskip\noindent\textbf{Healthcare systems.} A reliable, secure, and fast mobile communication can strengthen medical services, \textit{e}.\textit{g}., frequent data transfer from patients' body to the cloud or health care centers. Therefore, the relevant and urgent medical services could be predicted and delivered to the patients very fast.

\smallskip\noindent\textbf{Logistics and tracking.} The future mobile communication would also assist in inventory or package tracking using location based information systems. The most popular way would be to embed a radio frequency identification (RFID) tag and to provide a continuous connectivity irrespective of the geographic locations.

\smallskip\noindent\textbf{Industrial usages.} The zero latency property of 5G networks would help robots, sensors, drones, mobile devices, users, and data collector devices to have real-time data without any delay, which would help to manage and operate industrial functions quickly while preserving energy.
%\end{itemize}

\section{Real Demonstrations and Testbeds for 5G Networks}
\label{sec:Real Demonstrations and Test-beds for 5G Networks}
In this section, we present some real demonstrations and testbeds for 5G Networks. DOCOMO~\cite{docomo} is developing a real-time simulator for evaluating and simulating mMIMO, small-cell, and mmWave. The simulator has shown 1000-times increase in the system capacity, and 90\% users achieved 1Gbps data rate. DOCOMO also performed a real experiment in the year 2012, where data was uploaded at the speed of 10Gbps.

Samsung performed data transmission experiment using mmWave at 28 GHz frequency band and achieved the world's first highest data rate of 1.2Gbps on a vehicle running at the speed of $>$ 100km/h. Further, when the vehicle was nearby a stop, the data transmission speed was achieved up to 7.5Gbps. In the experiment, the peak data rate was more than 30-times faster as compared to the state-of-the-art 4G technology. Samsung is also developing array antennas that have nearly zero-footprint and reconfigurable antenna modes~\cite{samsung}. Ericsson has achieved the speed of 5Gbps in a demonstration.\footnote{{\scriptsize \url{http://www.ericsson.com/research-blog/5g/ericsson-research-hits-5gbps-5g-labs-demo/}}} Huawei at University of Surrey in Guildford is developing a testbed, which would be used for developing methodologies, validating them, and verifying a C-RAN for an ultra-dense network.\footnote{{\scriptsize\url{http://www.fiercewireless.com/tech/story/huawei-invests-5g-test-bed-university-surrey/2014-11-04}}}

The European Commission funded METIS~\cite{metis} has developed more than 140 technical components, including: air interface technologies, new waveforms, multiple access and MAC schemes, multi-antenna and mMIMO technologies, multi-hop communications, interference management, resource allocation schemes, mobility management, robustness enhancements, context aware approaches, D2D communication, dynamic reconfiguration enablers, and spectrum management technologies. They have also implemented some testbeds, as: three D2D communication related testbeds, one massive machine communications related testbed, and one related to waveform design. There are other projects funded by the European Commission working on the development of 5G networks, as: 5GNOW (\url{http://www.5gnow.eu/}), TROPIC (\url{www.ict-tropic.eu}), MCN (\url{www.mobile-cloud-networking.eu}), COMBO (\url{www.ict-combo.eu}), MOTO (\url{www.fp7-moto.eu}), and PHYLAWS (\url{www.phylaws-ict.org}).

\textit{Small-cells}: Sprint, Verizon, and AT\&T in the United States, Vodafone in Europe, and Softbank in Japan are developing femtocells ~\cite{DBLP:journals/wc/WangZ14}. Alcatel-Lucent, Huawei, and Nokia Siemens Networks has been involved in the development of plug-and-play SBSs~\cite{DBLP:journals/wc/NiCWL14}. \textit{mMIMO}: TitanMIMO by Nutaq is a testbed for 5G mMIMO. The TitanMIMO-4 testbed provides a realistic throughput by aggregating the entire mMIMO channel into a central baseband processing engine.\footnote{{\scriptsize\url{http://www.kanecomputing.co.uk/pdfs/Nutaq_TitanMIMO-4.pdf}}} Lund University and National Instruments, Austin are also involved in the development of testbeds for mMIMO~\cite{DBLP:conf/globecom/VieiraMNMKLWOET14}. DOCOMO~\cite{docomo} is developing a real-time simulator for mMIMO.

\textit{C-RANs}: China Mobile Research Institute has developed C-RANs~\cite{china}. The European Commission supported iJoin (\url{http://www.ict-ijoin.eu/}) is also involved in developing C-RAN architectures, especially, RANaaS~\cite{DBLP:journals/cm/RostBDGLMSW14}. IBM, Intel, Huawei, and ZTE are developing C-RANs~\cite{DBLP:journals/network/WuZHW15}. OpenAirInterface (OAI)~\cite{DBLP:journals/ccr/NikaeinMMDKB14} is an open experimentation and prototyping platform created by the Mobile Communications Department at EURECOM. OAI used for two purposes as: C-RANs and M2M communication. \textit{Full duplex radio}: groups at Stanford and Rice Universities are focusing on the development of full duplex radios~\cite{DBLP:conf/sigcomm/BharadiaMK13}. \textit{CRNs}: Vodafone Chair Mobile Communications Systems at the TU Dresden has created a testbed for studying CRNs in the future cellular networks~\cite{DBLP:conf/vtc/DannebergDF14}.\footnote{{\scriptsize\url{https://mns.ifn.et.tu-dresden.de/Chair/Pages/The-Chair.aspx}}} \textit{mmWave}: DOCOMO~\cite{docomo}, Huawei, the European Commission funded projects, and New York University~\cite{DBLP:journals/access/RappaportSMZAWWSSG13} are developing methodologies and testbeds for mmWave and have successfully performed various experiments.

\section{Concluding Remarks}
\label{sec:Concluding Remarks}
In this survey, we discussed salient features, requirements, applications, and challenges involved in the development of the fifth generation (5G) of cellular mobile communication that is expected to provide very high speed data transfer and ubiquitous connectivity among various types of devices. We reviewed some architectures for 5G networks based on the inclusion of small-cells, cognitive radio networks, device-to-device communication, and cloud-based radio access networks. We find out that energy consumption by the infrastructure is going to be a major concern in 5G networks, and hence, reviewed energy-efficient architectures. We figured out several open issues, which may drive the future inventions and research, in all the architectures.

The development of new architectures is not only a concern in 5G networks; there will be a need for handling other implementation issues in the context of users, \textit{e}.\textit{g}., interference removal, handoff management, QoS guarantee, channel accessing, and in the context of infrastructures, \textit{e}.\textit{g}., load balancing. During our illustration, we included several new techniques, \textit{e}.\textit{g}., full duplex radios, dense-deployment techniques, SIC, DUD, mmWave, mMIMO, and VLC. We also discussed the current trends in research industries and academia in the context of 5G networks based on real-testbeds and experiments for 5G networks.

We conclude our discussion with a resonating notion that the design of 5G infrastructure is still under progress. The most prominent issues are enlisted below, and to provide elegant solutions to these issues would contribute in early deployment as well as in long run growth of 5G networks.
\begin{itemize}[noitemsep,nolistsep,leftmargin=.5cm]
  \item The security and privacy of devices, infrastructures, communication, and data transfer is yet to be explored. We believe that the current solutions based on encryption would not suffice in the future due to a huge number of devices. Intuitively, a solution that would use an authenticated certificate may be feasible~\cite{nisha}.

  \item The development of network devices, infrastructures, and algorithms must be self-healing, self-configuring, and self-optimizing to preform dynamic operations as per the need, for example, dynamic load balancing, QoS guarantee, traffic management, and pooling of residual resources.

  \item The cloud computing is an attractive technology in the current trend for various applications. We have reviewed C-RANs; however, the current solutions do not consider the impact of virtualization for backhaul data transfer, the trust of the cloud, inter-cloud communication, ubiquitous service guarantee, and real-time performance guarantee with zero-latency. Thus, the development of C-RANs must address the big question that how much virtualization is good?

  \item The multi-RATs are attractive solutions to access different RATs. However, would it be possible for devices to use more than one RAT at an identical time for uplink and downlink channels? Further, the network densification must quantify that how much network density is good?

  \item The design, development, and usage of user devices, service-application models, and, especially, the network devices must be affordable to cater the needs of overwhelming users, service providers, and network providers.

  \item The zero latency is a primary concern in most of the real-time applications and services, especially, in the tactile Internet. However, all the existing architectures and implementations of 5G networks are far from achieving the zero latency. Therefore, it is utmost desired that the latest real-time and ultra-reliable network configurations must be improved to a latency free environment.
\end{itemize}

%\phantomsection
%\section{Acknowledgements}
\medskip\medskip
\noindent\textbf{\Large{Acknowledgements}}

\smallskip

\noindent Authors are thankful to Shlomi Dolev and anonymous referees for remarks that improved the quality of the paper.

%\bibliographystyle{abbrv}
%{\footnotesize \bibliography{5G-RelatedWork-new}}
%\bibliography{5G-RelatedWork-new}

\end{document}